\documentclass[9pt,twocolumn,twoside]{gsajnl}

\usepackage[utf8]{inputenc}
\usepackage{amssymb}% http://ctan.org/pkg/amssymb
\usepackage{pifont}% http://ctan.org/pkg/pifont
\usepackage[T1]{fontenc}
\usepackage{amsfonts}
\usepackage{amsmath}
\usepackage[super]{natbib}
\usepackage{hyperref}
\usepackage[capitalise]{cleveref}
\usepackage{placeins}
\usepackage{nicefrac}
\usepackage{pythonhighlight}
\lstnewenvironment{mypython}[1][]{\lstset{style=mypython,#1}}{}

\usepackage{graphicx}
\usepackage{booktabs}

\usepackage{caption}        % Subfigure packages
\usepackage{subcaption}     % Subfigure packages
\usepackage{multirow}
\usepackage{xcolor}

\usepackage{microtype}
\usepackage{contour}
\usepackage{adjustbox}

\setlist[enumerate]{itemsep=-1mm}

\newcommand{\absdiv}[1]{%
  \par\addvspace{.5\baselineskip}% adjust to suit
  \noindent\textbf{#1}\quad\ignorespaces
}

\renewenvironment{thebibliography}[1]{
  \begin{oldthebibliography}{#1}
    \setlength{\itemsep}{0em}
    \setlength{\parskip}{0em}
}
{
  \end{oldthebibliography}
}

\newcommand{\xmark}{\ding{55}}%

\title{SLOctolyzer: Fully automatic analysis toolkit for segmentation and feature extracting in scanning laser ophthalmoscopy images}
\runningtitle{SLOctolyzer: Fully Automatic SLO Image Analysis}
\runningauthor{Burke, et al.}

\author[1,2]{Jamie Burke}
\author[2]{Samuel Gibbon}
\author[3,4]{Justin Engelmann}
\author[2]{Adam Threlfall}
\author[3]{Ylenia Giarratano}
\author[5]{Charlene Hamid}
\author[1]{Stuart King}
\author[6,7]{Ian J.C. MacCormick}
\author[2,5,7]{Tom MacGillivray}

\affil[1]{School of Mathematics, University of Edinburgh, Edinburgh, UK}
\affil[2]{Robert O Curle Ophthalmology Suite, Institute for Regeneration and Repair, University of Edinburgh, UK}
\affil[3]{Centre for Medical Informatics, University of Edinburgh, Edinburgh, UK}
\affil[4]{School of Informatics, University of Edinburgh, Edinburgh, UK}
\affil[5]{Clinical Research Facility and Imaging, University of Edinburgh, Edinburgh, UK}
\affil[6]{Institute for Adaptive and Neural Computation, School of Informatics, University of Edinburgh, Edinburgh, UK}
\affil[7]{Centre for Clinical Brain Sciences, University of Edinburgh, Edinburgh, UK }

\correspondingauthoraffiliation[$\ast$]{Corresponding and lead author; \\ Email address: \url{Jamie.Burke@ed.ac.uk}}

\begin{abstract}
\absdiv{Purpose} 
The purpose of this study was to introduce SLOctolyzer: an open-source analysis toolkit for en face retinal vessels in infrared reflectance scanning laser ophthalmoscopy (SLO) images.

\absdiv{Methods}
SLOctolyzer includes two main modules: segmentation and measurement. The segmentation module uses deep learning methods to delineate retinal anatomy, and detects the fovea and optic disc, whereas the measurement module quantifies the complexity, density, tortuosity, and calibre of the segmented retinal vessels. We evaluated the segmentation module using unseen data and measured its reproducibility.
 
\absdiv{Results} 
SLOctolyzer's segmentation module performed well against unseen internal test data (Dice for all-vessels = 0.91; arteries = 0.84; veins = 0.85; optic disc = 0.94; and fovea = 0.88). External validation against severe retinal pathology showed decreased performance (Dice for arteries = 0.72; veins = 0.75; and optic disc = 0.90). SLOctolyzer had good reproducibility (mean difference for fractal dimension = -0.001; density = -0.0003; calibre = -0.32 microns; and tortuosity density = 0.001). SLOctolyzer can process a 768 $\times$ 768 pixel macula-centred SLO image in under 20 seconds and a disc-centred SLO image in under 30 seconds using a laptop CPU.

\absdiv{Conclusions}
To our knowledge, SLOctolyzer is the first open-source tool to convert raw SLO images into reproducible and clinically meaningful retinal vascular parameters. It requires no specialist knowledge or proprietary software, and allows manual correction of segmentations and re-computing of vascular metrics. SLOctolyzer is freely available at \url{https://github.com/jaburke166/SLOctolyzer}. 

\absdiv{Translational Relevance}
SLO images are captured simultaneous to optical coherence tomography (OCT), and we believe SLOctolyzer will be useful for extracting retinal vascular measurements from large OCT image sets and linking them to ocular or systemic diseases.

\end{abstract}

\begin{document}

\maketitle     

\section{Introduction}
The retina is a highly vascularised light-sensitive tissue at the back of the eye. The retinal microvasculature branches from the optic nerve head along the inner surface of the retina and is commonly imaged using colour fundus photography, which typically covers a 35- to 50-degree field of view around the posterior pole. 

Confocal near infrared reflectance scanning laser ophthalmoscopy (SLO) utilise long wavelengths of around 820nm and also captures an en face view of the superficial retinal vessels \cite{terasaki2021recent}. SLO images are typically captured simultaneous to optical coherence tomography (OCT) scans of the inner retinal layers (supplementary \cref{suppfig:heyex_demo}) acting as a localiser to position the OCT beam at the back of the eye accurately and typically images a restricted field of view of 30-degrees. Additionally, confocal imaging leverages laser-point raster scanning to prevent stray light interference during acquisition \cite{HILDEBRAND201363}, thus providing higher contrast visualisation of the en face retinal vasculature over regular colour fundus images. 

The interferometry of OCT imaging allows for micron-level axial depth visualisation on the cross-section which is independent of biometric factors like axial length and corneal or refractive properties of the eye \cite{salmon2018axial}. When these parameters are accounted for, the transverse plane can also be imaged with micron-level precision, enabling standardised physical measurements of the en face retinal vessels using microns-per-pixel conversion factors \cite{scoles2022inaccurate}. In contrast, colour fundus imaging typically relies on heuristic methods, such as optic disc area normalisation, to achieve standardised measurements \cite{schanner2023impact}. This suggests that SLO imaging could potentially provide more accurate and clinically relevant measurements of en face retinal vessels compared to traditional fundus imaging, given known biometric factors of the eye are accounted for. Nevertheless, despite the high level of detail captured by SLO images, they are often overlooked, as the primary focus of OCT imaging remains on visualising the cross-sectional retinal layers.

This improved visualisation, possibly increased accuracy in measurement and growing abundance from OCT capture make the SLO imaging modality an exciting frontier of en face retinal vessel analysis. Manual delineation of en face retinal vessels is prohibitively time-consuming, labour-intensive, and prone to inaccuracies. Consequently, extensive research has focused on segmenting retinal vessels, primarily in the colour fundus modality \cite{mahapatra2024review}, with less emphasis on SLO images.

At the time of writing, very few methods have been reported methods for retinal vessel analysis on SLO images \cite{xu2008retinal, pellegrini2014blood, kromer2016automated, meyer2017deep, hatamizadeh2022ravir}, and none for other major landmarks on the SLO image such as the fovea and optic disc. Additionally, none of these methods have been made available to the research community, which has the potential to impact the standardisation of SLO-derived retinal vascular metrics. 

Equipping automatic imaging methods with the tools for measuring downstream features for research and clinical use is an important element of any research pipeline \cite{perez2011vampire, zhou2022automorph, engelmann2022robust, burke2023open, engelmann2024choroidalyzer}. This allows the general researcher to use these methods in an end-to-end manner, and analyse their own retinal images without specialist training or background in image processing. Unfortunately, there is a lack of similar pipelines for SLO images.

Accordingly, we aimed to develop and release a fully automatic analysis toolkit, SLOctolyzer, for segmentation and measurement of retinal vessels as seen on SLO images. Here, we introduce and quantitatively validate the segmentation models, trained on datasets primarily related to systemic health, to detect the en face retinal vessels into arteries and veins, and detect the optic disc and foveal pit.

We highlight several contributions that SLOctolyzer provides to the research community:
\begin{itemize}
    \item SLOctolyzer is the first open-source pipeline for converting raw SLO images into clinically meaningful, retinal vascular parameters.
    \item SLOctolyzer's measurement module computes several summary measures of the retinal vessels, such as fractal dimension, vessel density, tortuosity density, vessel calibre and central retinal artery/vein equivalents. 
    \item SLOctolyzer works on macula- or optic disc-centred images which usually accompany OCT capture. 
    \item SLOctolyzer supports manual annotation to correct erroneous retinal vessel segmentation maps.
    \item SLOctolyzer is suitable for batch processing large amounts of SLO images, collating results for ease of segmentation quality inspection and downstream statistical analysis.
\end{itemize}

\section{Methods}

\begin{figure*}[tb]
    \centering
    \includegraphics[width=0.75\textwidth]{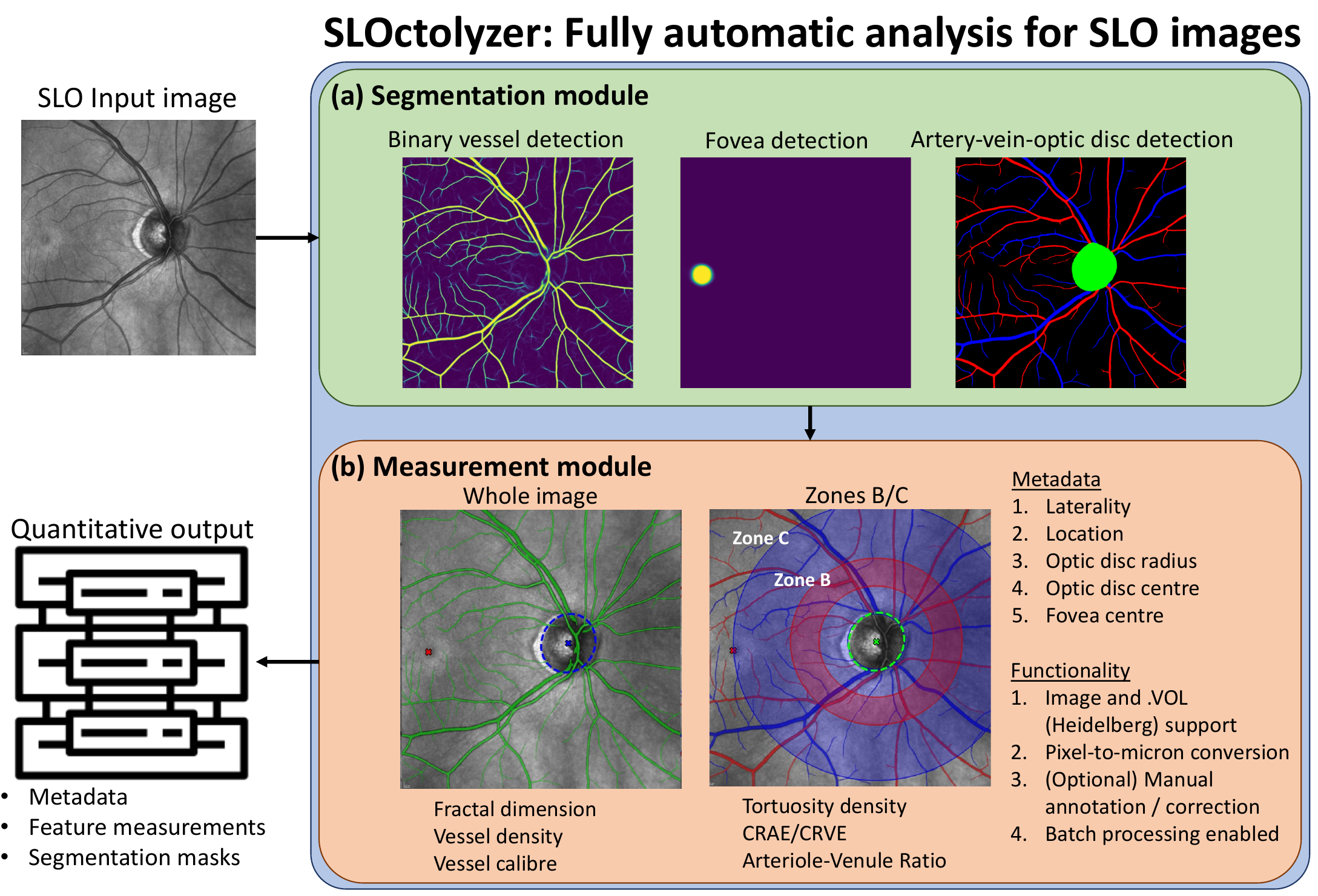}
    \caption{Summary of SLOctolyzer's analysis pipeline with a segmentation module (a) for binary vessel/artery/vein segmentation and fovea/optic disc detection, and a measurement module (b) for measuring retinal vascular parameters, with additional functionality for metadata extraction and manual annotation.}
    \label{fig:sloctolyzer_pipeline}
\end{figure*}

\cref{fig:sloctolyzer_pipeline} summarises the key elements of SLOctolyzer's pipeline, which contains a segmentation module and a measurement module (of which the latter is based on features and code produced by Automorph \cite{zhou2022automorph}). We have released SLOctolyzer in a frictionless manner so that it can be used without author permissions, proprietary software, or specialist training. SLOctolyzer is freely available here: \url{https://github.com/jaburke166/SLOctolyzer}.

\subsection{Data}
Data used to build the segmentation module include RAVIR \cite{hatamizadeh2022ravir}, a publicly available dataset of pathological SLO images with the arteries and veins segmented, one cohort of healthy volunteers and three cohorts related to systemic health: i-Test \cite{dhaun2014optical}, a cohort of pregnant women at late gestation who are experiencing either a normative, pre-eclamptic or fetal growth restricted pregnancy; PREVENT \cite{ritchie2012prevent, ritchie2013prevent}, a cohort of mid-life individuals, half of whom are at risk of developing later life dementia; FutureMS \cite{kearns2022futurems, chen2022longitudinal}, a cohort of individuals with newly diagnosed relapsing-remitting multiple sclerosis (MS). All studies and cohorts adhered to the Declaration of Helsinki and received relevant ethical approval, and informed consent from all subjects was obtained in all cases from the host institution.

\cref{tab:imagecharac_tab} summarises image characteristics of each cohort. All SLO images were captured using a spectral domain Heidelberg SPECTRALIS HRA+OCT imaging device (Heidelberg Engineering, Heidelberg, Germany) and were either macula-centred or optic disc-centred. SLO images from every study were grayscale and captured a field of view of 30-degrees at the posterior pole which approximately images a 9 mm$^2$ region of interest, using a broadband laser light source with wavelength 820nm. All cohorts used a HRA+OCT SPECTRALIS Standard Module except for the i-Test study, which used a portable version, the SPECTRALIS FLEX, for 24 participants out of the 93 available. Given a 30-degree field of view covering approximately a 9 mm$^2$ region of interest, the SLO images of these cohorts had either approximately 25 or 50 pixels per degree of visual angle, constituting a total imaging resolution of 768 $\times$ 768 or 1536 $\times$ 1536 (pixel height $\times$ width), which had transversal spatial sampling length-scales of approximately 11.71 microns-per-pixel or 5.85 microns-per-pixel, respectively.

\begin{table*}[tb]
\centering
\resizebox{\textwidth}{!}{%
\begin{tabular}{lllllllll}
\toprule
Study & Participants & Images & Right eyes & Retinal pathology & HRA+OCT Module & Image resolution [px] & Location \\
\midrule
RAVIR \cite{hatamizadeh2022ravir} & 23 & 23 & 14 & Yes & Standard & 768 $\times$ 768 & Disc \\
Healthy \cite{cameron2016modulation} & 7 & 7 & 7 & No & Standard & 1536 $\times$ 1536 & Disc \\
PREVENT \cite{ritchie2012prevent, ritchie2013prevent} & 144 & 285 & 142 & No & Standard & 1536 $\times$ 1536 & Disc \\
i-Test \cite{dhaun2014optical} & 93 & 186 & 93 & No & Standard \& FLEX & 768 $\times$ 768 & Macula \\
FutureMS \cite{kearns2022futurems, chen2022longitudinal} & 15 & 15 & 9 & No & Standard & 1536 $\times$ 1536 & Disc \\
\bottomrule
\end{tabular}}
\caption{Image characteristics of the five cohorts used to build SLOcolyzer's segmentation module. Image resolution is in pixels (for both lateral and axial directions), location refers to the centring of the scan, i.e. if it's macula-centred disc-centred. HRA, Heidelberg retina angiography; OCT, optical coherence tomography. px, pixels.}
\label{tab:imagecharac_tab} 
\end{table*}

SLOctolyzer's segmentation module contains three segmentation models, one for binary vessel detection, one for fovea detection and another for artery-vein and optic disc detection. \cref{tab:seg_to_cohort} outlines which cohorts and how many eyes/images of the cohorts each model used for training and evaluation. For all studies except for FutureMS, we selected all available participants for modelling. For FutureMS, a subset of 15 participants from the wider cohort \cite{chen2022longitudinal} were selected. These images were chosen based on image features relevant for vessel segmentation, such as blur, illumination, contrast, and abnormal features like vessel tortuosity or optic nerve head atrophy (supplementary \cref{suppfig:avod_dataset}). Similarly, for the i-Test cohort, 15 participants were selected from the wider cohort for building the artery-vein-optic disc (AVOD) model.

\begin{table}[tb]
\centering
\resizebox{0.5\textwidth}{!}{%
\scalebox{0.95}{\begin{tabular}{lllllll}
\toprule
\multirow{2}{*}{Model} & \multicolumn{5}{c}{Cohort (eyes/images)} & \multirow{2}{*}{Total} \\
\cmidrule(l){2-6}
 & RAVIR & Healthy & PREVENT & i-Test & FutureMS &   \\
 \midrule
Vessel & 23 & 7  & \xmark & \xmark & \xmark & 30 \\
Fovea & 23 & 7 & 285 & 186 & 15 & 516 \\
AVOD & \xmark & \xmark & \xmark & 15 & 15 & 30\\
\bottomrule
\end{tabular}}}
\caption{Number of eyes (images) used for each segmentation model stratified by cohort. AVOD, artery-vein-optic disc.}
\label{tab:seg_to_cohort} 
\end{table}

\subsection{Ground truth labels}

\subsubsection{Binary vessel detection} 
A total of 30 SLO images (RAVIR \& Healthy, \cref{tab:imagecharac_tab}) were used to train and evaluate the segmentation model for binary vessel detection. 23 of these were sourced from the RAVIR dataset, of which manual pixel-level annotation of the arteries and veins were done by the same, experienced retinal image grader \cite{hatamizadeh2022ravir}. Segmentation over the optic nerve head was only performed if the artery and vein classes were able to be resolved. Vessel labels were defined as both the artery and vein class combined. The remaining seven SLO images were made available from healthy eyes and were manually segmented by an experienced image analyst (author S.G.) using ITK-Snap \cite{py06nimg}, which supports pixel-level annotation.

\subsubsection{Fovea detection}
Fovea detection was performed by the same experienced grader (author J.B.) and used all 516 SLO images. 

The definitions of the fovea are guided by features on the image such as brightness and shape, rather than being biologically driven. We define the foveal pit as a single pixel coordinate on the en face SLO and cross-sectional OCT B-scan at the centre of the foveola centralis. The foveola is a circular zone of approximately 350-micron width which is avascular, creating a depression at the centre of the macula which pushes the inner retinal layers laterally \cite{FORRESTER20161}. On an OCT B-scan, the depression appears as a dip in the retina and is identified as a hyperreflective pixel at the centre of this dip, often aligning with a ridge formed at the photoreceptor layer. On the en face SLO, the depression appears as a dark spot (sometimes showing a hyperreflective dot in its centre) where smaller arterioles and venules, branching from major arteries and veins, converge to as they thin in calibre in the macula. See supplementary \cref{suppfig:SLO_fovea} and \cref{suppfig:OCT_fovea} and supporting text for exemplar en face and cross-sectional images with arrows indicating the detected foveal pit.

For the i-Test cohort, access to both the SLO and OCT was possible, which allowed the fovea coordinate on the SLO to be cross-referenced using it's transversal position on the corresponding fovea-centred OCT B-scan. The fovea location on each OCT B-scan was detected using Choroidalyzer \cite{engelmann2024choroidalyzer}, a fully automatic toolkit for choroid analysis and fovea detection in OCT images. For the remaining images, only the SLO was available and the fovea coordinate was selected manually, using an in-house GUI written in Python (version 3.11.6).

We assessed the accuracy of single-grader fovea detection through grader J.B. re-selecting the fovea pixel coordinate for all 516 images two months after the initial manual grading, the results of which can be found in supplementary \cref{supptab:fov_repeatability}. Repeatability against the initial i-Test fovea coordinates acted as comparison to ground truth, and the remaining images were used to assess intra-rater repeatability. Intra-rater repeatability had an average intra-class correlation (ICC(3,1)) of 0.99 for both $x$ and $y$ coordinates, and comparison to ground truth scored an average ICC of 0.82 across both $x$ and $y$ coordinates.  

The final fovea coordinate used for modelling was computed as the average of the two pixel coordinates selected.

\subsubsection{Artery-vein-optic disc detection}
Three experienced image analysts (authors J.B., S.G., A.T.) were each given 14 SLO images to manually annotate the arteries, veins, and optic disc. For each SLO image, a binary vessel map was generated using the binary vessel detection model. Each grader then used ITK-Snap \cite{py06nimg} for anatomical segmentation of arteries, veins and the optic disc. See supplementary \cref{suppfig:AV_class_rules} and \ref{suppfig:AV_protocol} for the vessel labelling rules and segmentation protocol each rater followed for consistent and comparative segmentation.

Each set of 14 images had two images with distinct overlap between pairwise graders. Thus, there were a total of 30 unique SLO images used for modelling: 15 from the i-Test cohort and 15 from the FutureMS cohort. This overlap allowed for a total of six SLO images which could be used for inter-rater assessment. We calculated inter-rater agreement for all-vessel, artery, vein, and optic disc segmentation using the Dice coefficient, whose results can be found in supplementary \cref{supptab:avod_interrater}. All values were found to be greater than 0.93, suggesting an excellent degree of consistency between graders.

We also had a clinical ophthalmologist (I.M.) qualitatively rate the artery and vein segmentations for all 30 SLO images. For each SLO image, we asked I.M. to rate its image quality and the segmentation quality of the artery and vein annotation using a 5-point ordinal scale between -2 (very bad) and 2 (very good). The results of this rating can be found in supplementary \cref{supptab:qualitative_segs}. 63\% (19/30) were rated as `very good', with the remaining annotations graded as `good', except for only one which was rated as `okay'. This example can be seen in supplementary \cref{suppfig:poorest_adjud}, which had two, small miss-labelled vessels. Importantly, there were no `bad' or `very bad' ratings for these manual annotations.

\subsubsection{The RAVIR dataset}
The RAVIR dataset \cite{hatamizadeh2022ravir} was intended to be used for training the artery-vein-optic disc (AVOD) model, as it contains suitable examples of severe retinal pathology. However, before modelling we reviewed the segmentation labels and observed inconsistent artery and vein classification, an example of which can be seen in supplementary \cref{suppfig:ravir_correction}. Therefore, we decided to use the RAVIR dataset as an external test set for the AVOD model, after correcting for any misclassified veins or arteries manually using ITK-Snap \cite{py06nimg}. This external test set presents a significant challenge to the AVOD model as it contains examples of severe retinal pathology.

Before evaluating the AVOD model, the corrections to the RAVIR dataset were reviewed by a clinical ophthalmologist (I.M.) against the original labels in a masked, randomised fashion. For each image, I.M. was presented with two segmentation labels without revealing their source. He was then asked to rate the SLO image quality and each artery-vein classification using a 5-point ordinal scale from -2 (very bad) to 2 (very good), and select which of the labels he preferred, with options for none or both. The results are shown in supplementary \cref{supptab:ravir_adjud}, and the ophthalmologist preferred over 64\% (14/23) of the corrected labels, and the labels for the remaining images were equally of good quality, according to I.M. Additionally, we segmented the optic disc for this external test set for comparison with our optic disc segmentation.

\subsection{Segmentation module}
% SLOctolyzer's first module contains three separate segmentation models with the following architectures and training pipelines.
During training, we applied the same data augmentations randomly to each model's training data. This included random horizontal and vertical flips ($P=0.5$), affine transformations to allow random rotation and scaling, uniformly sampled in $[0, 90]$ and $[0.4, 1.6]$, respectively ($P=1/3$). We also applied a Gaussian blur with random window size and standard deviation uniformly sampled in $[1, 20]$ and $[0.1, 4.0]$, respectively ($P=1/4$). Finally, we applied random brightness and contrast adjustments, altering the brightness and contrast using uniformly sampled scale factors $[0.8, 1.2]$ and $[0.75, 1.25]$, respectively.

We used python 3.12, PyTorch 2.0, Segmentation Models PyTorch \cite{Iakubovskii2019} and MONAI \cite{cardoso2022monai}.

\subsubsection{Binary vessel detection}
The model and training pipeline has been discussed previously \cite{threlfall2023publicly}. Briefly, the model was a custom UNet deep learning architecture trained from scratch, with details on how to reproduce the model architecture shown in supplementary \cref{suppfig:binary_Unet} with surrounding text.

The model was trained on a total of 30 SLO images. The dataset was split randomly into training (n=24), validation (n=2) and test (n=4) sets. Offline data augmentation randomly cropped each image into 20 patches of size 320 $\times$ 240 to artificially increase the training set, reduce computational load and accelerate the training rate.

The model was trained to minimise the Dice Focal loss \cite{cardoso2022monai}, using the Adam \cite{kingma2014adam} optimiser (with default optimiser hyperparameters) for 600 epochs using a batch size of 20. The initial learning rate was set as 1e-3 until epoch 300, at which point the learning rate was reduced to 1e-4 for the remainder of training. 

\subsubsection{Fovea detection}
Fovea segmentation is a significantly easier task than vessel detection or artery-vein classification, so our model used a pre-trained \cite{Iakubovskii2019}, lightweight UNet deep learning architecture with a MobileNetV3 \cite{howard2019searching} backbone, previously trained using ImageNet \cite{deng2009imagenet}. Details on how to reproduce the model architecture are shown in supplementary code listing \ref{suppcode:fovea_code} with surrounding text. 

This model used all available images from all cohorts, comprising 516 SLO images. The dataset was split into training ($n=359$), validation ($n=104$) and test sets ($n=54$) at the participant-level, so there was no overlap of participants between sets. All images were resized to a common image resolution of 768 $\times$ 768 pixels for training. To facilitate learning through segmentation, training labels were circular binary masks whose centre was the fovea coordinate with a radius of 60 pixels.

The model was trained to minimise the binary cross entropy loss using the AdamW \cite{loshchilov2017decoupled} optimiser for 200 epochs using a batch size of 8. During training, we applied two sets of random augmentations per image, per epoch, artificially doubling the batch size. An initial learning rate of 5e-4 was set, and after a 10-epoch linear warm up to a learning rate of 5e-3, it decayed following a cosine relationship for 90 epochs, with the learning rate set at 5e-4 for the remaining 100 epochs.

\subsubsection{Artery-vein-optic disc detection}
Given the number of segmentation tasks and dataset size, we similarly opted for fine-tuning a pre-trained UNet deep learning model, this time using a deep ResNet101 \cite{he2016deep} backbone, previously trained on ImageNet \cite{deng2009imagenet}. Details on how to reproduce the model architecture are shown in supplementary code listing \ref{suppcode:fovea_code} and surrounding text. 

The AVOD model was trained on a total of 30 SLO images. The dataset was split into training (n=20), validation (n=4) and test (n=6) sets, with equal proportion of cohorts (and thus macula- and disc-centred images) in each set. All images were resized to a common image resolution of 768 $\times$ 768 pixels for training.

The model was trained to minimise the Dice Focal loss \cite{cardoso2022monai}, using the Adam \cite{kingma2014adam} optimiser (with default optimiser hyperparameters) for 300 epochs using a batch size of 4, and a learning rate of 5e-4.

\subsection{Measure module}\label{subsec:features}
\begin{figure*}[tb]
    \centering
    \includegraphics[width=\textwidth]{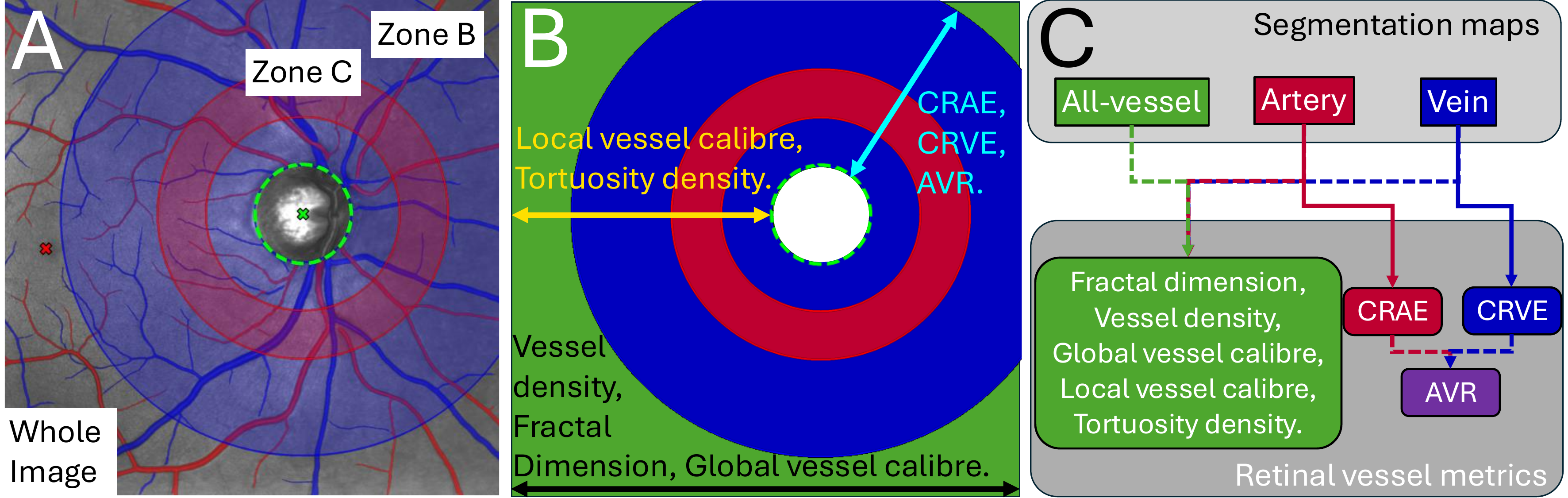}
    \caption{(A) Disc-centred SLO image with segmentations and regions of interest overlaid. (B) Region of interest masks for the whole image (green), zone C (blue) and zone B (red) with the names of vessel metric, colour coordinated with arrows indicating which regions they are measured in. (C) Flowchart of which vessel metrics are computed for which segmentation map.}
    \label{fig:sloctolyzer_metrics}
\end{figure*}

SLOctolyzer's second module is a suite of tools for measuring retinal vascular parameters and are based on the features measured when processing colour fundus images using Automorph \cite{zhou2022automorph}. \cref{fig:sloctolyzer_metrics} shows the different regions of interest defined and the measurements taken per region and per vessel map.

We define only one region of interest for macula-centred images and three regions of interest for disc-centred images to compute feature measurements on. The region of interest common across macula- and disc-centred images is the whole image (\cref{fig:sloctolyzer_metrics}(B), green). Relevant to only disc-centred images are the second and third regions of interest, which define two concentric rings centred at the optic disc called zones B and C \cite{cameron2016modulation}. Zone B takes an annulus from one-half diameter (D) away from the optic disc margin to a full diameter away (0.5D -- 1D) (\cref{fig:sloctolyzer_metrics}(B), red), and zone C measures from the optic disc margin to two diameters away (0D -- 2D) (\cref{fig:sloctolyzer_metrics}(B), blue). 

\cref{fig:sloctolyzer_metrics}(B) shows which metrics are measured across which regions of interest. Across the whole image, we measure global features of density, complexity and calibre, computed using vessel density (the ratio of vessel pixels to all pixels), fractal dimension (Minkowski-Bouligand dimension) \cite{falconer2004fractal} and global vessel calibre (ratio of vessel pixels to skeletonised vessel pixels) --- these measurements are not taken for localised zones B and C. For every region of interest (whole image and zones B and C), we measure local vessel calibre and tortuosity density \cite{grisan2003novel} across individual vessel segments whose endpoints are defined by arteriovenous crossing and bifurcations. Global and local vessel calibre are very similar metrics, but the latter provides a more granular approach to computing calibre from individual vessel segments --- see supplementary \cref{suppfig:global_vs_local_calibre} and supporting text.

\cref{fig:sloctolyzer_metrics}(C) shows a simple flowchart of which measurements are taken across which vessel map. All of these aforementioned features are measured across artery, vein and all-vessel maps (\cref{fig:sloctolyzer_metrics}(C), green). For artery and vein vessel maps only, we measure central retinal artery and vein equivalents (CRAE and CRVE) using the Knudston approach \cite{knudtson2003revised}, respectively, as well as the arteriole-to-venule (AVR) ratio (\cref{fig:sloctolyzer_metrics}(C), red-blue-purple). Note that although CRAE/CRVE/AVR are typically only measured for zones B and C in disc-centred images, they are also computed across the whole image and for macula-centred images too.  

\subsection{SLOctolyzer's pipeline}
\cref{fig:sloctolyzer_pipeline} describes the analysis pipeline for SLOctolyzer. Please see supplementary \cref{suppfig:sloctolyzer_input} and \cref{suppfig:sloctolyzer_output} with supporting text for more information on SLOctolyzer’s setup, interface and output.

\subsubsection{Input}
SLOctolyzer supports macula-centred and disc-centred SLO images inputted as regular image files (\verb|.jpeg|, \verb|.png|, etc.), as well as \verb|.vol| file formats (from Heidelberg Engineering imaging devices). In the former, the end-user can \textit{optionally} input information on spatial sampling, laterality (right/left) and location (macula-/disc-centred). If the laterality and location of the SLO image is not inputted, these are inferred based on the detected locations of the fovea and optic disc after segmentation, and measurements are outputted in pixel space. For inputting \verb|.vol| files, these store the image data and metadata necessary for converting measurements from pixel space to physical space --- other file formats such as \verb|.e2e| and \verb|.dcm| are currently not supported, but we are working on this presently. A configuration file is also provided for the end-user to specify input/output directories, whether the pipeline skips over files which have unexpected errors during processing (helpful for large-scale batch processing), and the option to ignore saving out segmentation maps (to reduce memory consumption during processing).

\subsubsection{Segmentation \& measurement}
During segmentation, the image is first resized to a common image resolution of 768 $\times$ 768 pixels as the segmentation module was trained on this image resolution, thus making the pipeline invariant to different spatial sampling length-scales (assuming a fixed field of view). Moreover, this resizing prevents any segmentation error in images with larger image resolution due to the central artery/vein light reflex \cite{threlfall2023publicly} which appears as a bright strip in the centre of a vessel, potentially leading to a large vessel being perceived as two, smaller and parallel vessels. Upon resizing, full segmentation of the vasculature, optic disc and fovea is performed and the probability segmentation maps are then resized to the image's original image resolution. After thresholding with a value of 0.5 to produce binary masks, post-processing is applied to remove small false positive regions, and disconnected vessels are joined to their neighbouring vessels to remove small gaps. 

The foveal pit is defined as the centroid of the fovea binary mask. If the SLO image is disc-centred, the optic disc binary mask is modelled as an ellipse, and it's centre and diameter are the ellipse's centre and the average of its major and minor axis lengths, respectively. If the SLO image is macula-centred, measurements on the all-vessel/artery/vein maps are only taken across the whole image. If the SLO image is disc-centred, measurements are taken across the whole image and for zones B and C, which are defined using the optic disc diameter. 

\subsubsection{Output}
Measurements will be saved out per file, alongside a process log and key metadata stored located in the file's metadata or inferred during analysis (laterality, location, fovea centre, optic disc diameter and centre). Optionally, segmentation masks will be saved out, as well as helpful visualisations to inspect segmentation quality. If running on a batch of images, a collated \verb|.xlsx| file will be saved out row-wise compiling all measurements and metadata from all image files analysed. Finally, a folder of visualisations, superimposing the segmentations onto the SLO image will be saved out per file which will aid segmentation quality checking.

While we do not have an automatic way to assess segmentation quality, an additional feature of SLOctolyzer is its ability to correct manual segmentations. Given a saved-out segmentation mask of the arteries, veins and optic disc, the end-user may correct any erroneous areas using ITK-Snap \cite{py06nimg} and save the corrected version in the same folder as the original mask. Upon rerunning the pipeline again, it will use this corrected annotation and re-compute the measurements for that image.

\subsection{Statistical analysis}

Each segmentation model, after model selection based on the performance against their respective validation set, was evaluated on their respective, internal test set. Additionally, for the AVOD model, we used the RAVIR  dataset \cite{hatamizadeh2022ravir} as an external test to measure the segmentation module’s robustness to severe retinal pathology.  

We used standard segmentation metrics such as area under the receiver operating characteristic curve (AUC) and the dice similarity coefficient (at a standard threshold of 0.5) to evaluate segmentation performance. We also used mean absolute error (MAE) to measure performance on downstream, retinal vascular parameters fractal dimension and local vessel calibre (in microns [$\mu$m], where transversal spatial sampling length-scales were available and pixels [px] otherwise). We used MAE to measure the error in predicted fovea pixel coordinate (along both axes) and optic disc area.

For internal test set evaluation, the whole image was used as the region of interest due to the low sample size of disc-centred images in each internal test set. For external test set evaluation, all regions of interest (whole image, zone B and zone C) were evaluated, as RAVIR contains only disc-centred images. 

We also measured the reproducibility of SLOctolyzer on downstream retinal vascular parameters, using the first 60 participants (120 eyes) of the i-Test cohort (whose data was available at the time of analysis). SLO images were captured simultaneous to macula-centred, OCT volume capture when enhanced depth imaging mode was toggled on and off, creating a repeated, albeit unregistered, pair of SLO images per eye. We measured all retinal vascular parameters previously mentioned across the whole image, and report mean absolute error and Pearson (P), Spearman (S) and intra-class correlation (ICC(3,1)) coefficients for population-based reproducibility analysis. We also report Bland-Altman plots \cite{bland1986statistical} to assess the distribution of residuals.

Additionally, we report the reproducibility of SLOctolyzer at the eye-level using a measure of individual-level measurement noise $\lambda$ \cite{engelmann2024applicability}. We express the variability of each measured feature within an eye in units of the feature's overall population variability. $\lambda$ is computed per feature, per eye, as the ratio between the standard deviation of within-eye measurements and the standard deviation of between-eye measurements, across the population. For convenience, $\lambda$ is presented as a percentage, with 0 as the optimal value. Note that the population are all white women with a mean (SD) age of 34.7 (5.2) years, at mean (SD) gestation of 35.6 (3.4) weeks, and thus we do not expect there to be huge variation across the population.

Finally, we measured the execution time of the SLOctolyzer's segmentation module and the whole pipeline on macula- and disc-centred images of different image resolutions. We ran each experiment 100 times and report the mean and standard deviation execution time in seconds. Timed experiments were run on a Windows laptop with a 4-year-old Intel Core i5 (8$^{\text{th}}$ generation) CPU and 16 Gb of RAM. For brevity, we will refer to this as the `laptop CPU' in the rest of the text.

\section{Results}

\begin{table*}[tb]
\centering
\resizebox{\textwidth}{!}{%

\begin{tabular}{@{}ccccccccccc@{}}
\toprule
  \multicolumn{1}{c}{\multirow{2}{*}{Metric}} &
  \multicolumn{2}{c}{Vessel} &
  \multicolumn{2}{c}{Fovea} &
  \multicolumn{2}{c}{Artery} &
  \multicolumn{2}{c}{Vein} &
  \multicolumn{2}{c}{Optic disc}\\
  \cmidrule(l){2-3}\cmidrule(l){4-5}\cmidrule(l){6-7}\cmidrule(l){8-9}\cmidrule(l){10-11}
  \multicolumn{1}{c}{} &
  \multicolumn{1}{c}{AUC} &
  \multicolumn{1}{c}{Dice} &
  \multicolumn{1}{c}{AUC} &
  \multicolumn{1}{c}{Dice} &
  \multicolumn{1}{c}{AUC} &
  \multicolumn{1}{c}{Dice} &
  \multicolumn{1}{c}{AUC} &
  \multicolumn{1}{c}{Dice} &
  \multicolumn{1}{c}{AUC} &
  \multicolumn{1}{c}{Dice}\\
  \midrule
 Segmentation & 0.99 & 0.91 & 0.99 & 0.88 &  0.99 & 0.84 & 0.99 & 0.85 & 0.99 & 0.94   \\
 \bottomrule
&  &  &  &  &  &  &  &  & & \\
\toprule
\multicolumn{1}{c}{} &
  \multicolumn{1}{c}{FD} &
  \multicolumn{1}{c}{Calibre [px]} &
  \multicolumn{1}{c}{C$_x$ [px]} &
  \multicolumn{1}{c}{C$_y$ [px]} &
  \multicolumn{1}{c}{FD} &
  \multicolumn{1}{c}{Calibre [px]} &
  \multicolumn{1}{c}{FD} &
  \multicolumn{1}{c}{Calibre [px]} &
  \multicolumn{2}{c}{Area [px$^2$]} \\
 \midrule
 Feature & 0.01 & 0.15 & 6.28 & 4.78 & 0.02 & 0.21 & 0.01 & 0.15 & \multicolumn{2}{c}{638.17} \\
 \bottomrule
\end{tabular}%
}
\caption{Segmentation results for each segmentation task (top). Mean absolute error across relevant measurements for each segmentation task (bottom). Calibre, local vessel calibre; FD, fractal dimension; C$_x$/C$_y$, mean absolute error in the horizontal/vertical component of the centroid for the fovea; px, pixels.}
\label{tab:seg_results} 
\end{table*}

\begin{figure*}[tb]
    \centering
    \includegraphics[width=\textwidth]{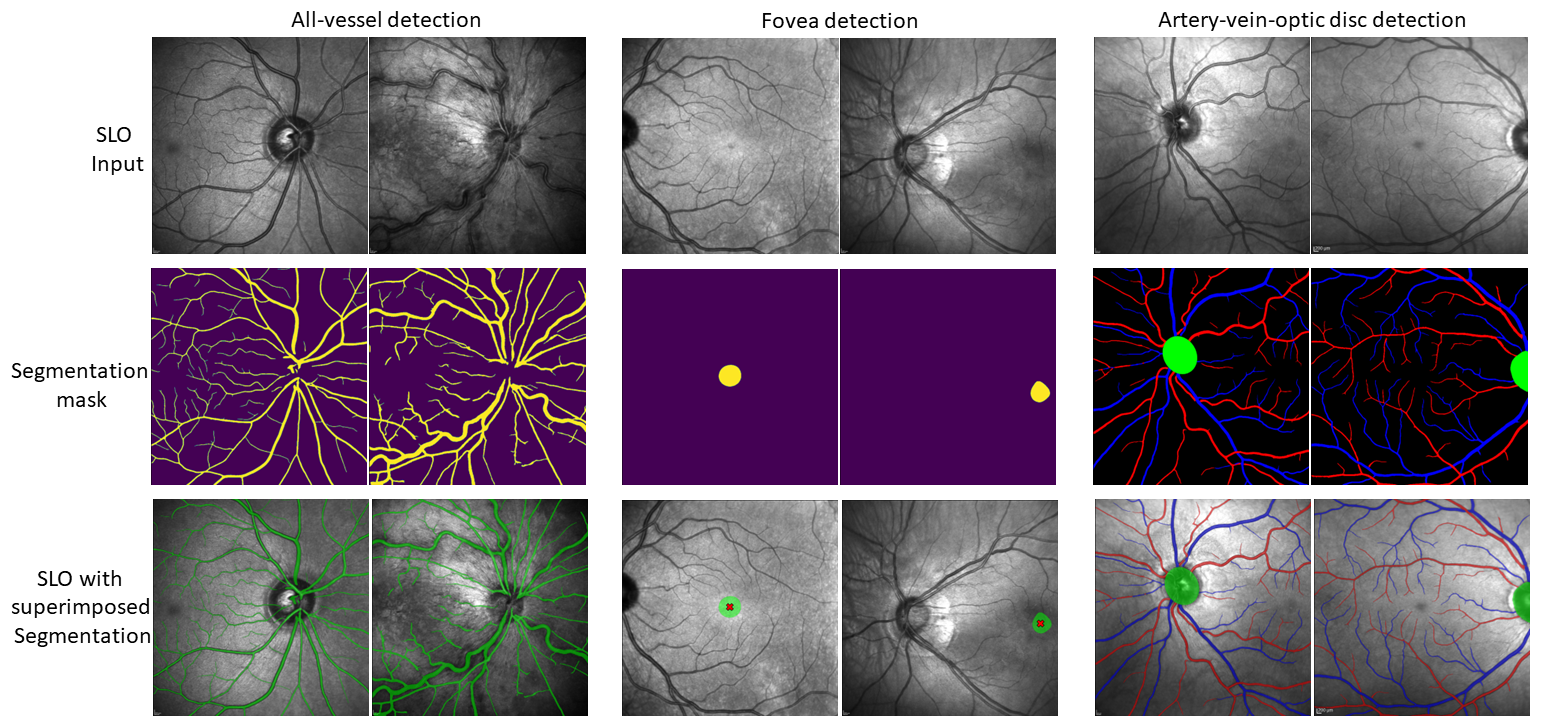}
    \caption{Predicted segmentation masks from each model for two examples randomly selected from their internal test sets.}
    \label{fig:sloctolyzer_internal_examples}
\end{figure*}

\subsection{Evaluation}
Segmentation results are presented in \cref{tab:seg_results} for all three models for their respective internal test sets. All models had a strong AUC score (all greater than 0.98), suggesting the raw probability maps were well calibrated in being able to score highly the relevant pixels related to each of their segmentation tasks, with the fovea and optic disc detection performing best, which was to be expected because these are easier segmentation tasks than vessel detection. Upon thresholding, both binary vessel and AVOD models scored high dice scores for all-vessel detection (binary vessel model: 0.91, AVOD model: 0.86), and the AVOD model scored slightly lower dice scores for artery and vein classification (artery, 0.84; vein, 0.85). This was unsurprising, because distinguishing between artery and vein is significantly more challenging than differentiating retinal vasculature from retinal tissue. \cref{fig:sloctolyzer_internal_examples} shows example segmentation outputs from the internal test sets.

The MAE for fovea coordinate detection was 6.3 and 4.8 pixels for the horizontal and vertical axes (see \cref{tab:seg_results}), thus reporting an MAE of 7.9 pixels diagonally. Images had an image resolution of 768 $\times$ 768 pixels, covering approximately a 9 mm$^2$ region of interest, which makes their transversal spatial length-scale approximately 11.71 microns-per-pixel in both axes. Thus, this diagonal MAE corresponds to approximately 92.7 microns, which is well within the estimated 350-micron width of the foveola centralis \cite{FORRESTER20161}.

Similarly, an MAE of 638 square pixels for optic disc area corresponds to approximately an MAE of 0.08 mm$^2$, which is not clinically significant given the wide distribution of optic disc sizes \cite{hoffmann2007optic}, the observed change across the myopic spectrum \cite{mishra2022assessment} and even differences across different devices \cite{brautaset2016repeatability}. MAE for local vessel calibre was around 0.2 pixels, which corresponds to approximately 2.32 microns, and is similarly an insignificant error given population distributions for arterioles and venules using confocal SLO imaging \cite{garg2022normative}.

\begin{figure*}[tb]
    \centering
    \includegraphics[width=\textwidth]{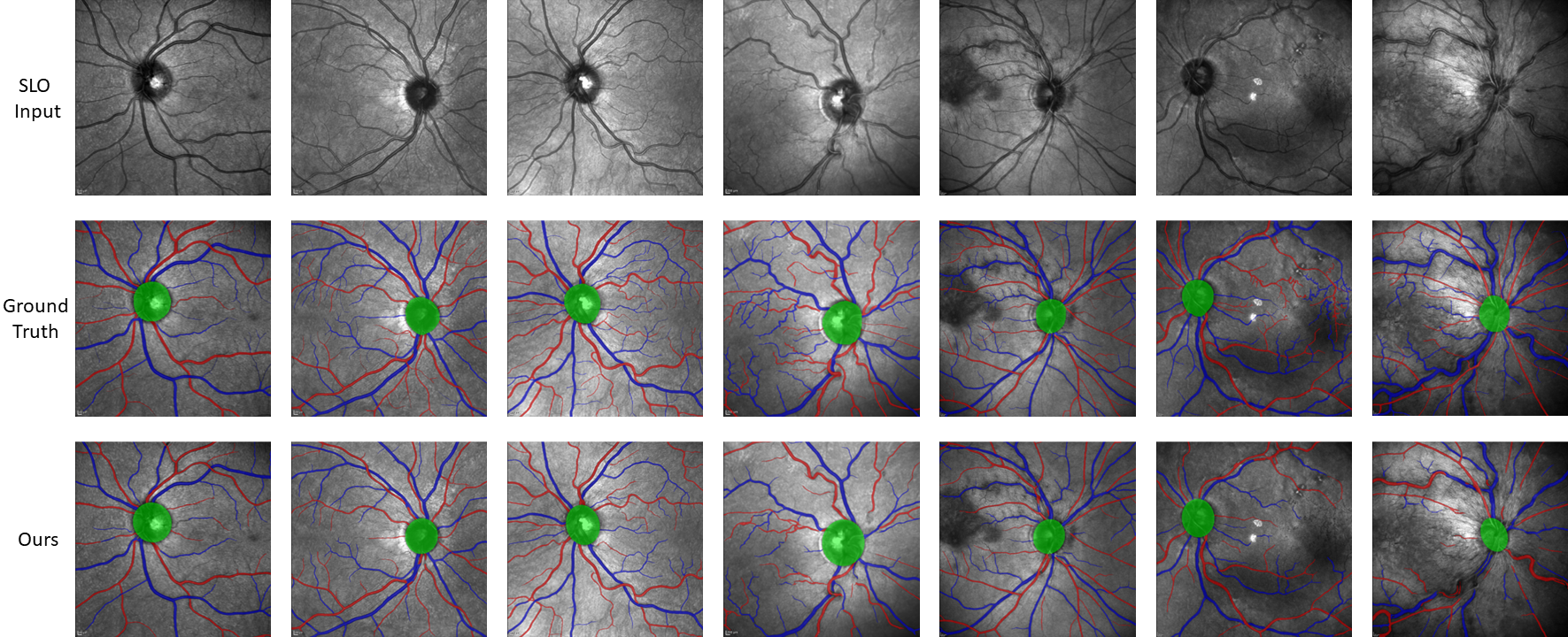}
    \caption{Visual comparison between ground truth and predicted segmentations for the external test set, RAVIR, for artery-vein-optic disc detection. The images were selected to show progressing eye pathology from left to right.}
    \label{fig:sloctolyzer_external_examples}
\end{figure*}

The results of our segmentation module evaluated against the external test set is shown in \cref{tab:ravir_results}. The RAVIR dataset \cite{hatamizadeh2022ravir} presents a significant challenge due to retinal pathology heavily featuring in the 23 SLO images. The model performs worse compared with the unseen, internal test data which relate to systemic health (dice coefficient averaged across zones: artery, 0.72; vein, 0.76; optic disc, 0.90), but quantitatively performs well for those images with mild pathology. Given the segmentation models were trained on images with grossly normal retinae related to systemic disease, it is perhaps no surprise that the performance of the model decreases as the extent of retinal pathology increases. \cref{fig:sloctolyzer_external_examples} qualitatively compares the ground truth labels (middle row) with SLOctolyzer's predictions (bottom row), for SLO images of increasing retinal pathology from left-to-right. While the major vessels are classified correctly for the majority of cases, as the pathology becomes more severe we see smaller and more tortuous vessels ignored, and segmentations of the major vessels misclassified.

\begin{table*}[tb]
\centering
\resizebox{\textwidth}{!}{%
\begin{tabular}{@{}lcccccccccccc@{}}
\toprule
 \multicolumn{1}{c}{\multirow{2}{2cm}{Region of interest}} &
  \multicolumn{4}{c}{Artery} &
  \multicolumn{4}{c}{Vein} &
  \multicolumn{4}{c}{Optic disc}\\
  \cmidrule(l){2-5}\cmidrule(l){6-9}\cmidrule(l){10-13}
  \multicolumn{1}{c}{} &
  \multicolumn{1}{c}{AUC} &
  \multicolumn{1}{c}{Dice} &
  \multicolumn{1}{c}{FD} &
  \multicolumn{1}{c}{Calibre [px]} &
  \multicolumn{1}{c}{AUC} &
  \multicolumn{1}{c}{Dice} &
  \multicolumn{1}{c}{FD} &
  \multicolumn{1}{c}{Calibre [px]} &
  \multicolumn{1}{c}{AUC} &
  \multicolumn{1}{c}{Dice} &
  \multicolumn{2}{c}{Area [px$^2$]} \\
  \midrule
 Whole Image & 0.94 & 0.72 & 0.04 & 0.69 & 0.96 & 0.75 & 0.07 & 0.98 & 0.97 & 0.90 & \multicolumn{2}{c}{788.50} \\
 Zone B & 0.99 & 0.73 & 0.04 & 1.01 & 0.99 & 0.77 & 0.06 & 1.18 & - & - & \multicolumn{2}{c}{-}\\
 Zone C & 0.96 & 0.72 & 0.04 & 0.79 & 0.97 & 0.76 & 0.05 & 1.15 & - & - & \multicolumn{2}{c}{-}\\
 \bottomrule
\end{tabular}

}
\caption{Segmentation and feature measurement mean absolute errors of the artery-vein-optic disc (AVOD) segmentation model against the RAVIR external test set. Calibre, local vessel calibre; AUC, area under the receiver operating characteristic curve; FD, fractal dimension; px, pixels.}
\label{tab:ravir_results} 
\end{table*}

\subsection{Reproducibility}
\cref{tab:sloctolyzer_repr} presents correlation values for all retinal vascular parameters for the binary vessel segmentation model (all-vessel), and the artery-vein-optic disc model. All correlations showed strong or excellent agreement among features, with fractal dimension and vessel density scoring the highest. Across the cohort, veins appeared as the poorest performing, but not by any significant margin.

\begin{table*}[tb]
\centering
\resizebox{\textwidth}{!}{%
\begin{tabular}{llllllllll}
\toprule
\multirow{2}{*}{Metric} & \multicolumn{3}{c}{All-vessel} & \multicolumn{3}{c}{Artery} & \multicolumn{3}{c}{Vein} \\
\cmidrule(l){2-4}\cmidrule(l){5-7}\cmidrule(l){8-10}
 & MAE & P/S & ICC(3,1) & MAE & P/S & ICC(3,1) & MAE & P/S & ICC(3,1) \\
\midrule
Fractal dimension & 0.008 & 0.93/0.92 & 0.97 (0.95, 0.98) & 0.005 & 0.95/0.94 & 0.97 (0.96, 0.98) & 0.007 & 0.89/0.89 & 0.94 (0.92, 0.96) \\
Vessel density & 0.001 & 0.95/0.95 & 0.97 (0.96, 0.98) & 0.002 & 0.94/0.94 & 0.97 (0.95, 0.98) & 0.001 & 0.94/0.92 & 0.97 (0.95, 0.98) \\
Global calibre {[}$\mu$m{]} & 1.58 & 0.88/0.89 & 0.93 (0.90, 0.95) & 1.60 & 0.84/0.83 & 0.91 (0.87, 0.94) & 1.61 & 0.87/0.87 & 0.93 (0.90, 0.95) \\
Local calibre {[}$\mu$m{]} & 1.50 & 0.87/0.88 & 0.93 (0.89, 0.95) & 1.47 & 0.85/0.85 & 0.92 (0.88, 0.94) & 1.41 & 0.87/0.87 & 0.93 (0.90, 0.95) \\
Tortuosity density & 0.03 & 0.74/0.75 & 0.85 (0.79, 0.9) & 0.01 & 0.80/0.80 & 0.89 (0.84, 0.92) & 0.02 & 0.70/0.68 & 0.83 (0.75, 0.88) \\
AVR/CRAE/CRVE {[}$\mu$m{]} & 0.03 & 0.77/0.79 & 0.87 (0.82, 0.91) & 6.28 & 0.85/0.86 & 0.92 (0.88, 0.94) & 10.10 & 0.79/0.80 & 0.89 (0.83, 0.92) \\
\bottomrule
\end{tabular}%
}
\caption{Population-based reproducibility performance of SLOctolyzer, reporting mean absolute error (MAE), Pearson (P), Spearman (S) and intra-class correlations (ICC(3,1)). All Pearson and Spearman correlations were statistically significant with P-values $P < 0.0001$. For readability, AVR/CRAE/CRVE were combined into one row --- AVR is dimensionless and specified in the all-vessel column, while CRAE and CRVE are in microns and only report for the artery and vein maps, respectively.}
\label{tab:sloctolyzer_repr} 
\end{table*}

\begin{figure*}[tb]
    \centering
    \includegraphics[width=\linewidth]{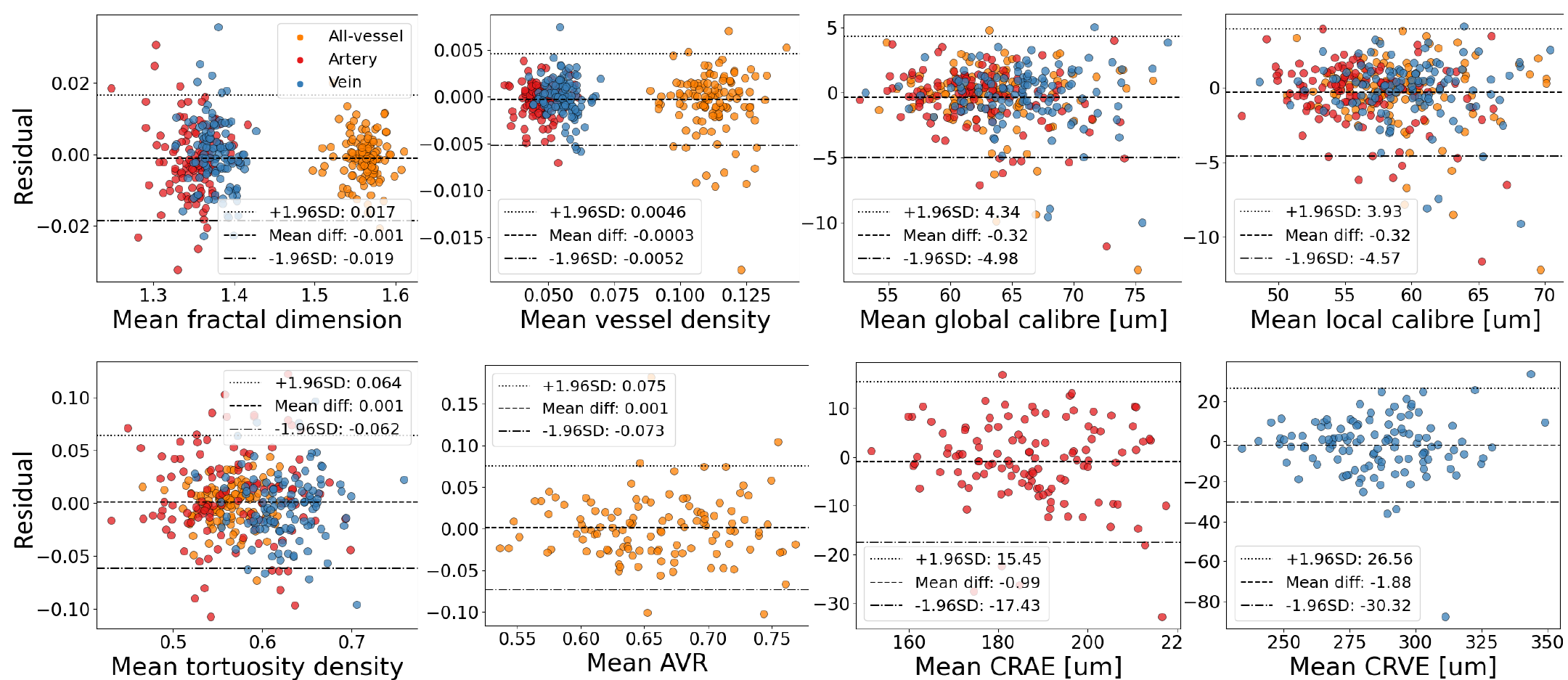}
    \caption{Bland-Altman plots of residual distributions from assessing the reproducibility of SLOctolyzer's segmentation module across all retinal vessel metrics. Scatter-plot residuals for arteries (red), veins (blue) and all-vessel (orange) are overlaid together.}
    \label{fig:sloctolyzer_repeatability}
\end{figure*}

\cref{fig:sloctolyzer_repeatability} shows Bland-Altman plots of the residuals between paired, repeated SLO images. We found that SLOctolyzer was highly reproducible, with the distribution of residuals according to most metrics centred around 0, showing no apparent trend. For fractal dimension, there was a mean difference (MD) of -0.001 with limits of agreement (LoA) of [-0.019, 0.017]; for vessel density there was a MD of -0.0003 and LoA of [-0.0052, 0.0046]; for global vessel calibre there was a MD of -0.32 $\mu$m and LoA of [-4.98, 4.34]; for local vessel calibre there was a MD of -0.32 $\mu$m and LoA of [-4.57, 3.93]; for tortuosity density there was a MD of 0.001 and LoA of [-0.062, 0.064]; for CRAE there was a MD of -0.99 $\mu$m and LoA of [-17.43, 15.45]; for CRVE there was a MD of -1.88 $\mu$m and LoA of [-30.32, 26.56]; finally, for AVR there was a MD of 0.001 and LoA of [-0.073, 0.075].

\begin{figure*}[tb]
    \centering
    \includegraphics[width=\textwidth]{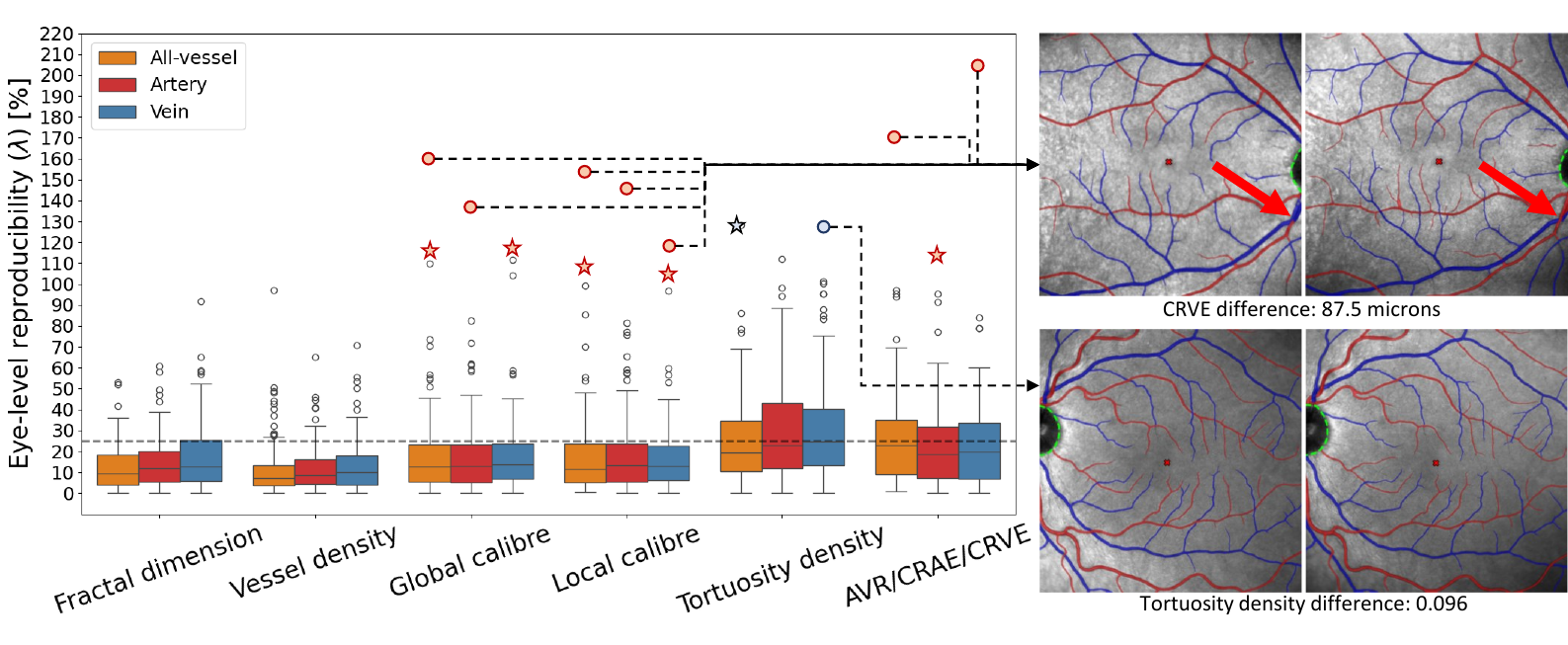}
    \caption{Reproducibility at the eye-level of each of the features outputted by SLOctolyzer. Box-plot distributions are shown for each feature, with all-vessels in orange, arteries in red and veins in blue. Note that the AVR/CRAE/CRVE features are presented together as CRVE/CRAE are measured across arteries (red) and veins (blue) separately, while AVR is a combination of the two, thus represented in the all-vessel class (orange). A horizontal, dashed line at the $\lambda$=25\% mark is shown as a visual aid. We also show the eyes (with segmentations overlaid) which had the greatest eye-level reproducibility measurement error for large-vessel calibre and vein tortuosity density (circular scatter-points). Eyes corresponding to the star-shaped scatter-points are shown in supplementary \cref{suppfig:poor_repeatability}.}
    \label{fig:ind_level_repr}
\end{figure*}

\cref{fig:ind_level_repr} reports the eye-level reproducibility of each of the features measured by SLOctolyzer. For fractal dimension, vessel density and global/local calibre, the upper, or third quartile is below 25\% across all vessel types. Thus, the range of measurement error for these features in the majority of the paired, but unregistered, eyes were below 25\% of the overall populations' variability. Given the homogeneous nature of the population's demographics, this is very reasonable. Moreover, vessel density has the smallest interquartile range among features (mean all-vessels, $\lambda = 11.7\%$; arteries, $\lambda = 12.1\%$; veins, $\lambda = 13.3\%$), suggesting this feature, closely followed by fractal dimension, had the least measurement noise. Tortuosity density and AVR/CRAE/CRVE appear as the most sensitive features, with veins as the most variable relative to arteries and all-vessels. 

The eyes with the worst reproducibility for vessel calibre and tortuosity density are shown to the right of the box plots, with dashed lines linking the major outlier scatter-points to the corresponding eye's SLO image. For all calibre metrics, the worst metric came from the same eye. This was due to an arteriovenous crossing near the edge of the image, where a major artery and vessel overlapped, and SLOctolyzer classified artery and vein differently (red arrows). This led to a large absolute difference of 87.5 microns in CRVE. For tortuosity density, the vessel maps do not look qualitatively different, but the absolute difference was 0.096. We believe that this is due to tortuosity density's sensitivity to arteriovenous crossings which disconnect the respective artery and vein vessel maps. Star-shaped scatter-points represent second to largest outliers for vessel calibre and tortuosity density and are shown in supplementary \cref{suppfig:poor_repeatability}, and are similarly due to arteriovenous crossings or vessel disconnectedness.

\subsection{Execution time}
\cref{tab:sloctolyzer_times} shows the execution time of SLOctolyzer's segmentation module and pipeline. Because the segmentation module resizes all images to a common image resolution of 768 $\times$ 768 pixels, the segmentation execution time across different locations and image resolutions are approximately equal. For binary vessel segmentation, it takes approximately 4.5 s/image, for the fovea detection model it takes approximately 0.6 s/image, and for AVOD detection it takes approximately 3.0 s/image. Thus, SLOctolyzer's segmentation module using the laptop CPU takes approximately 11.1 $\pm$ 0.5 s/image. Measuring the execution time of the entire pipeline on the laptop CPU took 18.9 $\pm$ 1.0s for a macula-centred SLO with image resolution 768 $\times$ 768 pixels, 28.6 $\pm$ 1.0s for a disc-centred SLO with image resolution 768 $\times$ 768 pixels and 111.0 $\pm$ 3.5s for a disc-centred SLO with image resolution 1536 $\times$ 1536 pixels. Disc-centred SLO images are more time intensive as measurements are made across the whole image, and for zones B and C across the all-vessel, artery and vein segmentation maps separately.

\begin{table*}[tb]
\centering
{\small
\scalebox{0.95}{\begin{tabular}{@{}lllll@{}}
\toprule
  \multicolumn{1}{c}{\multirow{2}{*}{Image type}} &
  \multicolumn{3}{c}{Segmentation module} &
  \multicolumn{1}{c}{\multirow{2}{*}{Whole pipeline}} \\
  \cmidrule(l){2-4}
  & Binary vessel & Fovea & AVOD &  \\
  \midrule
 Macula-centred ($768 \times 768$) & 4.8 (0.407) & 0.6 (0.067) &  2.9 (0.408) & \textbf{18.9 (0.984)} \\
 Disc-centred ($768 \times 768$) & 4.3 (0.147) & 0.6 (0.047) & 2.9 (0.440) & \textbf{28.6 (1.040)} \\
 Disc-centred ($1536 \times 1536$) & 4.2 (0.191) & 0.6 (0.051) & 2.7 (0.133) & \textbf{111.0 (3.540)} \\
 \bottomrule
\end{tabular}}%
}
\caption{Average (standard deviation) execution time in seconds of the segmentation module and SLOctolyzer's entire pipeline for SLO images of different location and image resolution.}
\label{tab:sloctolyzer_times} 
\end{table*}

\section{Discussion}

We have developed a fully automatic method, SLOctolyzer, for processing macula-centred and disc-centred SLO images which typically accompany OCT data but are often discarded in research. SLOctolyzer converts a raw SLO image into reproducible and clinically meaningful retinal vascular parameters, including fractal dimension, tortuosity, and vessel calibre. 

SLOctolyzer performed well on unseen data, but we observed a performance drop against an external test set featuring severe retinal pathology. SLOctolyzer's segmentation module was trained on datasets primarily related to systemic health and was thus designed for modelling the retinal vasculature in these disease contexts, so there was understandably a noticeable decrease in performance as the subjective extent of pathology increased. However, we believe SLOctolyzer's additional functionality to correct any erroneous segmentation maps and re-compute measurements will prove very useful for cases of severe retinal pathology. Regardless, we envision SLOctolyzer being a very useful tool for SLO image analysis in systemic disease, particularly as the nascent field of oculomics increases rapidly \cite{wagner2020insights} and the populations being studied often have grossly normal retinae.

SLOctolyzer's segmentation module had good reproducibility, with the residuals across all metrics for vessels, arteries and veins exhibiting no pattern and being centred around 0. In particular, the residuals for measurements of global and local calibre (in microns) appear to show proportionately small error in comparison to what is generally considered as clinically significant. Meta-analyses have found vessel calibre difference of 10 -- 15 microns between mid- and old-age, with a 20 micron increase in venule calibre associated with increased risk of stroke by 1.15 \cite{ikram2013retinal}.

The distributions of measurement error at the eye-level for fractal dimension, vessel density and local/global calibre were found to be reasonable given the homogeneous nature of the i-Test cohort studied for reproducibility --- with tortuosity density and large-vessel calibre having higher variability. We observed that small differences in the segmentation maps can have an inordinate impact on tortuosity density (supplementary \cref{suppfig:poor_repeatability}(A)). This highlights the apparent sensitivity of tortuosity density, regardless of whether the segmentation looks qualitatively different. Segmentation error does contribute to this error, as arteriovenous crossings result in artery-vein map disconnectedness which can have an inordinate impact on generating individual vessel segments to measure tortuosity on. This observation adds to a potential problem surrounding standardisation of retinal vascular parameters \cite{mcgrory2018towards}. 

For large-vessel calibre, the largest outliers came from the same eye, which showed a major arteriovenous crossing at the edge of the image causing an artery-vein miss-classification. Moreover, large-vessel calibre metrics like AVR/CRAE/CRVE are typically reserved for disc-centred SLO images as these show the major vessels in full as they branch from the optic nerve head. Thus, the unregistered nature of the macula-centred SLO pairs, combined with major vessels potentially cropped toward the edge of the image, likely contributed to the measurement error (supplementary \cref{suppfig:poor_repeatability}(B)). We hope in the future we may also assess SLOctolyzer's reproducibility for paired, disc-centred SLO images. 

On a 4-year-old Intel Core i5 (8$^{\text{th}}$ generation) laptop CPU, SLOctolyzer can segment an SLO image in around 11 seconds. For an SLO with image resolution 768 $\times$ 768 pixels, the entire pipeline takes 19 seconds for macula-centred images and 29 seconds for disc-centred images. It can be used interactively or for batch processing (supplementary \cref{suppfig:sloctolyzer_input}), converting raw SLO images into reproducible vascular metrics and outputting segmentation masks for quality inspection (supplementary \cref{suppfig:sloctolyzer_output}). 

There were some limitations associated with this work. Firstly, the SLO images used to build SLOctolyzer's segmentation module were sourced from only one manufacturer (Heidelberg Engineering) and we do not know SLOctolyzer's performance on SLO images from other devices. Moreover, our datasets only contained macula- and disc-centred SLO images captured at a field of view of 30-degrees, and we did not explicitly test the pipeline's performance at different locations or fields of view in the posterior pole.

Furthermore, most of the images were from individuals recruited to studies on systemic disease which did not have obvious retinal pathology and we therefore expect the segmentation module to not be entirely robust to these cases. Finally, these individuals were of white ethnicity, and so we do not know the pipeline's performance across varying retinal pigmentation. However, SLOctolyzer’s additional functionality allows for any unexpected segmentation errors to be corrected manually via ITK-Snap \cite{py06nimg}.

A technical limitation of the work is that SLOctolyzer's segmentation module is made up of three distinct models. The rationale for this was clear at the time of model construction: we could leverage an initial binary vessel segmentation model to help us manually annotate the arteries and veins for building an artery-vein detection model, a much more challenging task, with the fovea detection model utilising all available images due to the speed of ground truth labelling. However, these models now provide a quick and consistent way to annotate large datasets which, after inspection and correction by expert graders, could be used to build a much more robust segmentation model. 

Thus, we intend to improve SLOctolyzer's segmentation module in future work by constructing a single model to perform all segmentation tasks, and include a larger and more diverse dataset on which it will be trained and evaluated. These will include examples related to retinal pathology, capturing different regions and fields of view of the retina, and from multiple imaging devices and manufacturers.

We anticipate SLOctolyzer's relevancy for processing 820nm SLO images which traditionally accompany OCT image capture using Heidelberg Engineering imaging devices. However, further exploration will be conducted on its generalisability to other SLO imaging modalities, such as Optos ultra-wide field SLO \cite{sodhi2021feasibility}, or Heidelberg's more recent multicolour SLO \cite{zhang2020multicolor}.

\section{Conclusion}
We have developed a fully automatic method for extracting useful information on the retinal vessels from SLO images, including fractal dimension, tortuosity, and vessel calibre. SLOctolyzer, the first open-source analysis toolkit for macula- and disc-centred SLO images, addresses the lack of open-source tools for this image modality. It can leverage existing OCT datasets to provide reproducible and clinically meaningful retinal vascular parameters from the SLO images which are captured in parallel to OCT acquisition, but are often overlooked. Available on GitHub, SLOctolyzer requires no specialist training or proprietary software and supports batch processing and manual correction. We hope this tool will encourage the research community to reconsider the value of SLO images, facilitate collaboration and help to standardise retinal vascular parameters in SLO images.

\section*{Acknowledgements}
J.B. was supported by the Medical Research Council (grant MR/N013166/1) as part of the Doctoral Training Programme in Precision Medicine at the Usher Institute, University of Edinburgh. For the purpose of open access, the authors have applied a creative commons attribution (CC BY) licence to any author accepted manuscript version arising.

Data collection for the Retinal Imaging sub study in PREVENT Dementia was supported in part by the Alzheimer's Drug Discovery Foundation (project no. GDAPB-201808-2016196, ``Delivering novel neuro-retinal biomarkers for the early diagnosis of Alzheimer's disease''); NHS Lothian R\&D; British Heart Foundation Centre for Research Excellence Award III (RE/18/5/34216); a University of Edinburgh Innovation Initiative Grant award; a ARUK Scotland Network Centre Equipment award; the Edinburgh \& Lothians Health Foundation. Data collection for i-Test was funded by the Wellcome Leap In Utero programme. These funding sources were not involved in designing, conducting, or submitting this work. 

\section{Conflicts of Interest}
The authors declare no conflicts of interest.

\bibliographystyle{unsrt}
\bibliography{references}

\onecolumn
 
\setcounter{figure}{0}
\renewcommand{\thefigure}{S\arabic{figure}}
\setcounter{table}{0}
\renewcommand{\thetable}{S\arabic{table}}
\pagebreak

\section*{SUPPLEMENTARY MATERIALS}

\section*{Simultaneous OCT + SLO capture during acquisition}
\begin{figure}[H]
\centering
\includegraphics[width=0.7\textwidth]{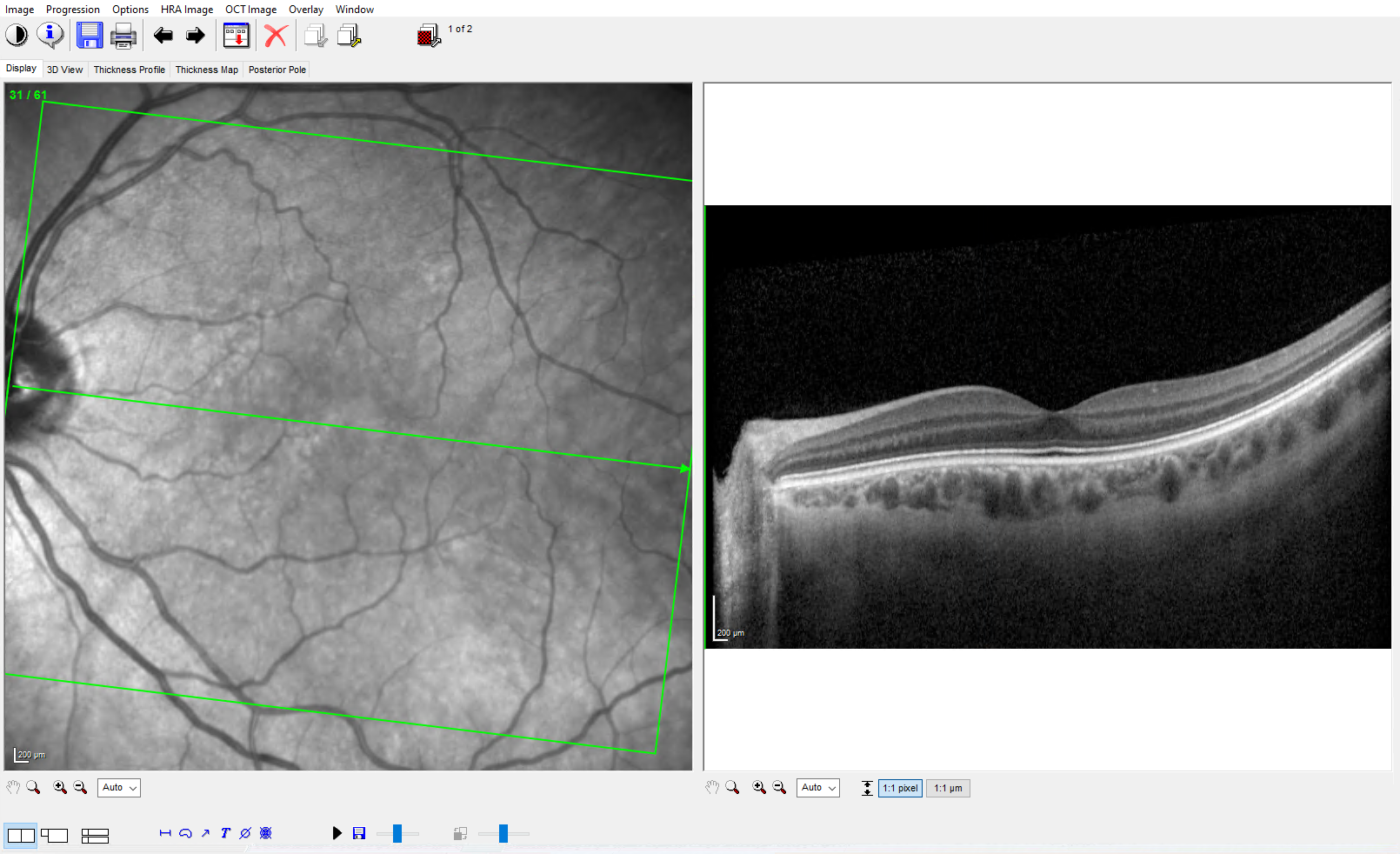}
\caption{Screenshot from Heidelberg Eye Explorer (HEYEX) of an OCT volume capture during acquisition. (Left) the confocal SLO image with the acquisition location of the OCT volume overlaid in green. (Right) the corresponding fovea-centred OCT B-scan of the OCT volume.}
\label{suppfig:heyex_demo} 
\end{figure}

Supplementary \cref{suppfig:heyex_demo} shows a screenshot from the Heidelberg Eye Explorer (HEYEX) software (version 1.12.1.0) (Heidelberg Engineering, Heidelberg, Germany) of an OCT volume for an individual’s left eye. During OCT capture, the confocal SLO image (left) is used to reference the location of the B-scans (right) captured during acquisition. This is advantageous particularly for registering follow-up scans. 

\section*{SLOctolyzer's training data for artery-vein-optic disc detection}

\begin{figure}[H]
\centering
\includegraphics[width=0.75\textwidth]{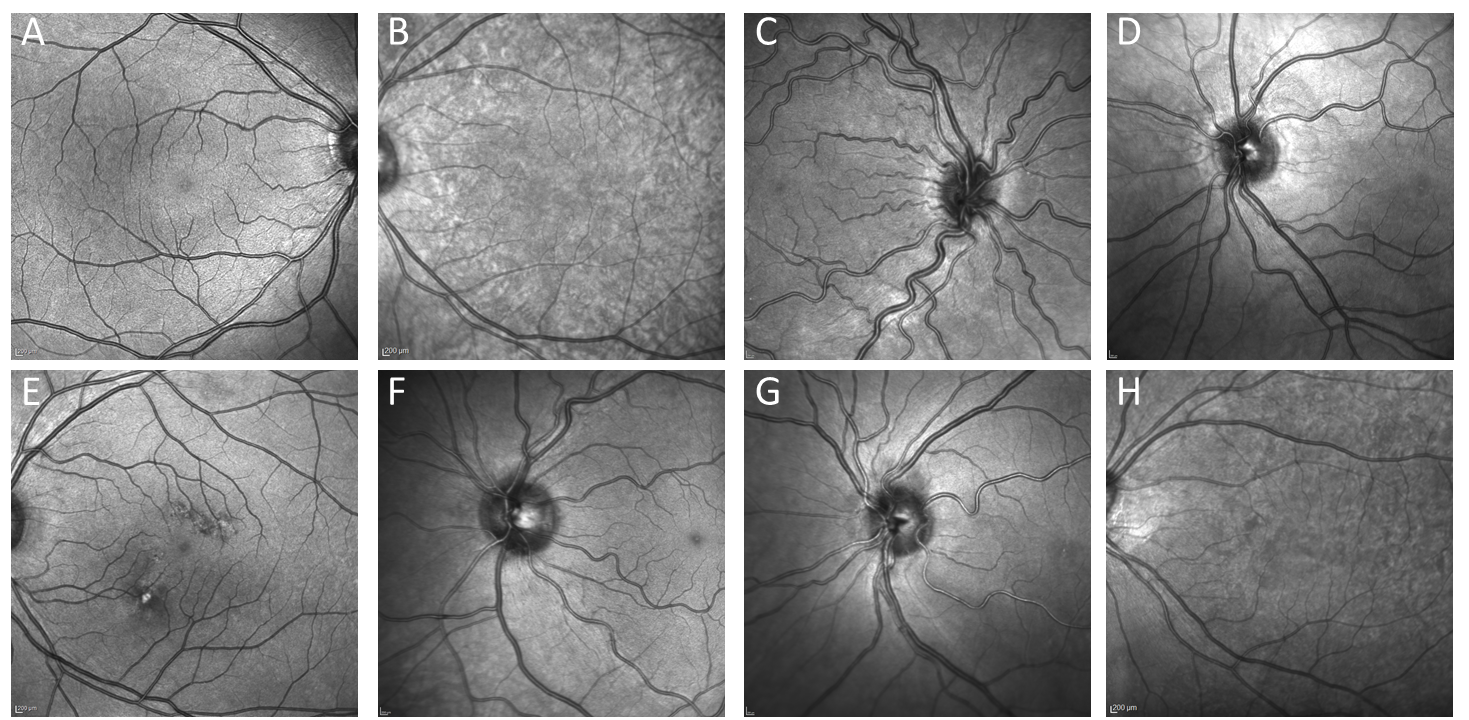}
\caption{Selection of the data used to build the artery-vein-optic disc segmentation model, highlighting features of interest in the image which are advantageous for diverse training, such as blur (A), contrast artefacts (B,E,H), non-uniform illumination (D, F, G), vessel tortuosity (C) and optic nerve head atrophy (H).}
\label{suppfig:avod_dataset} 
\end{figure}

Supplementary \cref{suppfig:avod_dataset} shows some exemplary SLO images selected from the i-Test and FutureMS cohorts for building the artery-vein-optic disc segmentation model. The total set of 30 SLO images were selected to provide a variety of image-features during training and evaluation, such as blur, non-uniform illumination, and contrast artefacts, as well as abnormal retinal features such as vessel tortuosity or optic nerve head atrophy.

\section*{Manual detection of the fovea}
\subsection*{Examples of detecting the fovea}
As part of the manual detection of the fovea coordinate on SLO images, author J.B. manually selected the fovea coordinate for all cohorts but the i-Test one. 

\begin{figure}[H]
\centering
\includegraphics[width=0.7\textwidth]{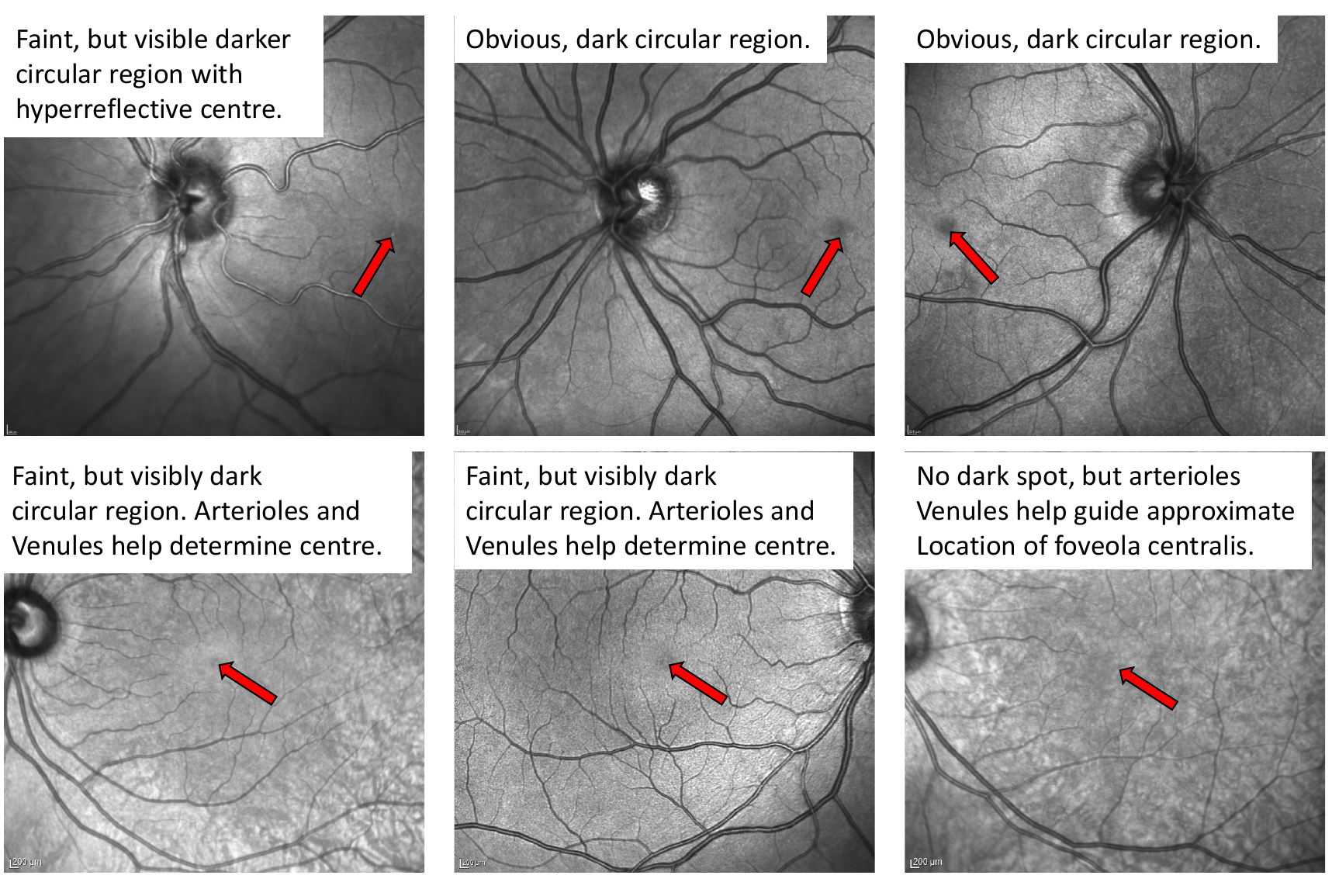}
\caption{Examples of  foveal pit detection for en face SLO images, with text overlaid to describe the detection process in each case.}
\label{suppfig:SLO_fovea} 
\end{figure}

For en face SLO fovea detection the pixel coordinate representing the fovea pit was detected as the centre of a small, dark and circular region representing the depression on the retinal tissue (sometimes showing a hyperreflective spot at the centre of this region, representing the foveal pit). This is typically seen in the centre of the macula-centred SLO image, and at the centre-row but skewed to the far right- or left-column for disc-centred images, depending on the eye's laterality. To aid detection, the small arterioles and venules branching from the major vessels were used to guide the centre of the macula and thus the fovea pit for more challenging cases. Supplementary \cref{suppfig:SLO_fovea} shows some examples of en face SLO images with their foveal pit identified by red arrows, with supporting text to describe the detection in each case.

\begin{figure}[H]
\centering
\includegraphics[width=0.7\textwidth]{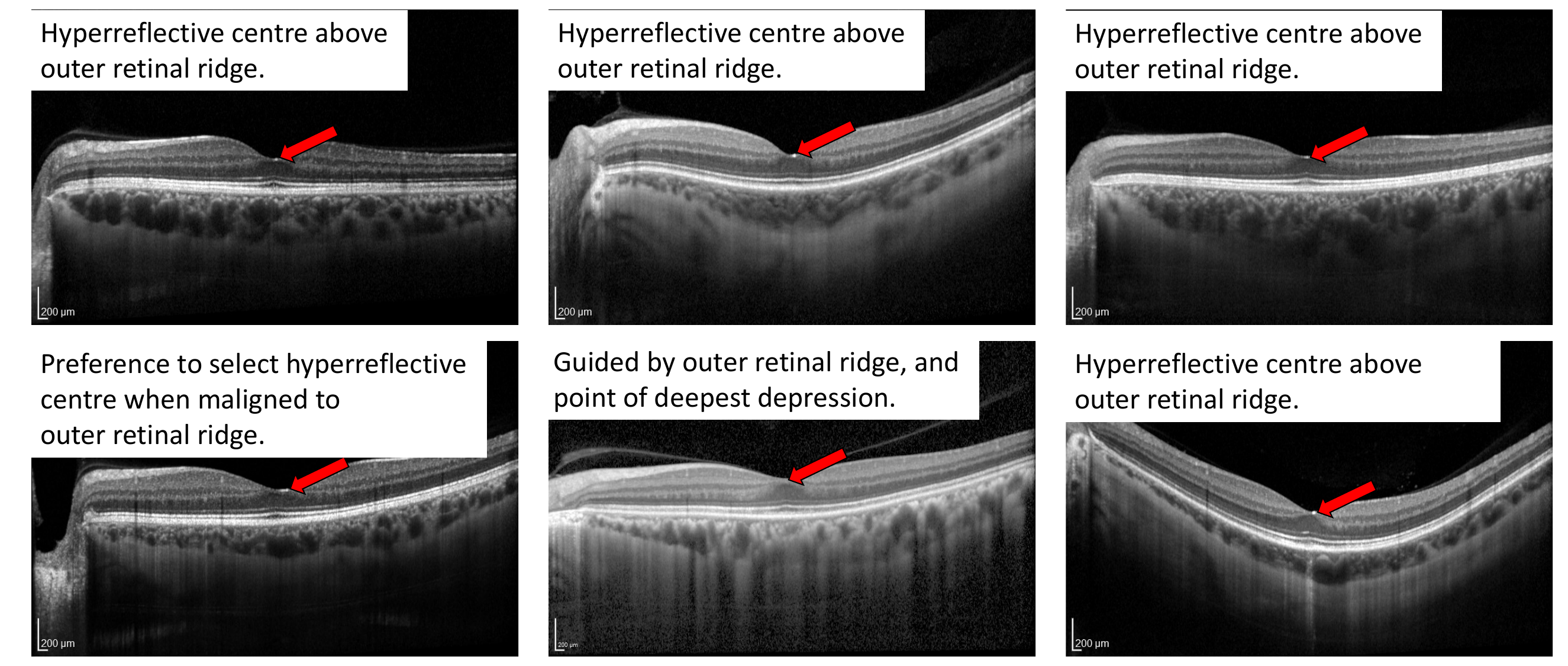}
\caption{Examples of foveal pit detection for fovea-centred, cross-sectional OCT B-scans, with text overlaid to describe the detection process in each case.}
\label{suppfig:OCT_fovea} 
\end{figure}

For a fovea-centred OCT B-scan, we used Choroidalyzer \cite{engelmann2024choroidalyzer} to define the foveal pit as a single pixel coordinate is at the point of deepest depression in the B-scan. This appears as a dip in the retina with a small hyperreflective region in the centre of this depression, often aligned with a ridge formed at the photoreceptor layer. Fortunately, this detection process is deterministic through Choroidalyzer, and was cross-referenced to the corresponding en face SLO image as we know the transversal position of the fovea on the B-scan and the exact location of acquisition of the B-scan on the SLO image. Supplementary \cref{suppfig:OCT_fovea} shows some examples of cross-sectional OCT B-scan with their foveal pit identified by red arrows, with supporting text to describe the detection in each case.

\subsection*{Rater repeatability}
As a measure of rater repeatability, author J.B. re-selected the fovea coordinate for all 516 SLO images two months after the initial manual detection, and compared the $(x,y)$-positioning using mean absolute error and intra-class correlation. Like the training pipeline for the fovea detection model, fovea segmentation masks were generated for each batch of coordinates and compared using the area under the receiver operating characteristic curve (AUC) and dice coefficient. Evaluation was performed after all images were resized to a common image resolution of 768 $\times$ 768 pixels for interpretable comparison.

The repeatability results are listed below in supplementary \cref{supptab:fov_repeatability}. Results for the i-Test cohort represent inter-rater repeatability (manual detection against accurate ground truth data from cross-referenced OCT B-scan using Choroidalyzer \cite{engelmann2024choroidalyzer}), and results for the remaining four cohorts of data represent intra-rater repeatability. Intra-rater repeatability had an excellent average intra-class correlation of 0.99 and a strong intra-class correlation for inter-rater repeatability of 0.82 was also reported.

\begin{table}[H]
\centering
{
\begin{tabular}{@{}lllllll@{}}
\toprule
  \multicolumn{1}{c}{} &
  \multicolumn{1}{c}{AUC} &
  \multicolumn{1}{c}{Dice} &
  \multicolumn{2}{c}{ICC(3,1)} &
  \multicolumn{2}{c}{MAE (px)}\\
  \cmidrule(l){4-5}\cmidrule(l){6-7}
  & & & x & y & x & y \\
  \midrule
 Others & 0.96 & 0.93 & 0.99 & 0.99 & 3.39 & 3.00 \\
 i-Test & 0.94 & 0.89 & 0.77 & 0.87 & 4.75 & 4.52 \\ \bottomrule
\end{tabular}%
}
\caption{Repeatability of detecting the fovea on the i-Test dataset (which measure inter-rater repeatability) and the remaining cohorts (which measure intra-rater repeatability). AUC, area under the receiver operating characteristic curve; ICC, intra-class correlation for single fixed raters; MAE, mean absolute error (in pixels)}
\label{supptab:fov_repeatability} 
\end{table}

\section*{Manual annotation of artery-vein-optic disc}\label{suppsec:annotate_AVOD}
The optic disc was classified as both the cup and rim together, and its margin was defined by the sharpest intensity transition between the darker rim and brighter retinal tissue constituting the retinal nerve fiber layer. For accurate artery-vein classification, the list below is a set of classification rules each grader followed:

\begin{enumerate}
    \item \textbf{Alternating} rule: It’s common to see arteries and veins appear on the fundus in an alternative nature, for efficient oxygen transport \cite{ishibazawa2019accuracy}.
    \item \textbf{Junction} rule: It is common to observe arteriovenous crossings (a junction of four lanes) and same-class vessel bifurcations (a junction of three lanes). Given a crossing where there are more than three lanes, it’s likely to be an arteriovenous crossing. Otherwise, it is a same-class vessel bifurcation.
    \item \textbf{Thickness} rule: Veins are typically thicker than arteries \cite{garg2022normative}. 
    \item \textbf{Brightness} rule: Veins are typically darker than arteries \cite{brinchmann1986intensity}.
\end{enumerate}

\begin{figure}[H]
\centering
\includegraphics[width=0.4\textwidth]{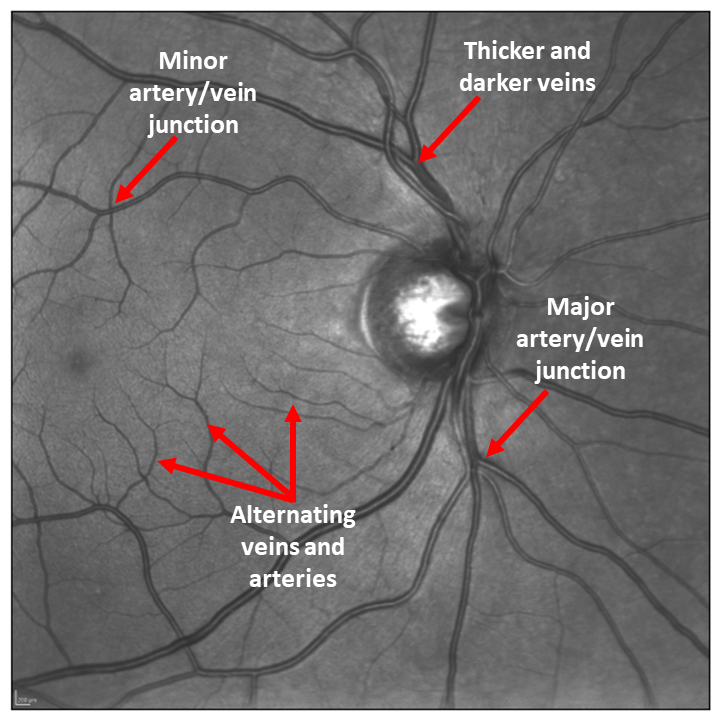}
\caption{An exemplary SLO image with annotations to describe each of the four rules for artery-vein classification.}
\label{suppfig:AV_class_rules} 
\end{figure}

Supplementary \cref{suppfig:AV_class_rules} shows an exemplary SLO image and annotations to describe each of the rules followed by the graders.

A protocol was created to ensure consistent and comparable segmentations by each grader using ITK-Snap \cite{py06nimg}. After application of the binary vessel segmentation model to detect all the vessels in each SLO image, the graders followed the rules listed below during pixel-level annotation:
\begin{enumerate}\setlength\itemsep{0em}
    \item Given an arteriovenous crossing, select the class which is most obviously overlaid on top the other class. If unsure, select the largest vessel.
    \item Do not classify any vessels within the optic disc.
    \item Complete any disconnected vessels.
    \item Do not extend vessels which have visibly not been segmented.
    \item Remove false positive pixels, i.e. retinal tissue which has been classed as vessel.
\end{enumerate}

\begin{figure}[H]
\centering
\includegraphics[width=0.75\textwidth]{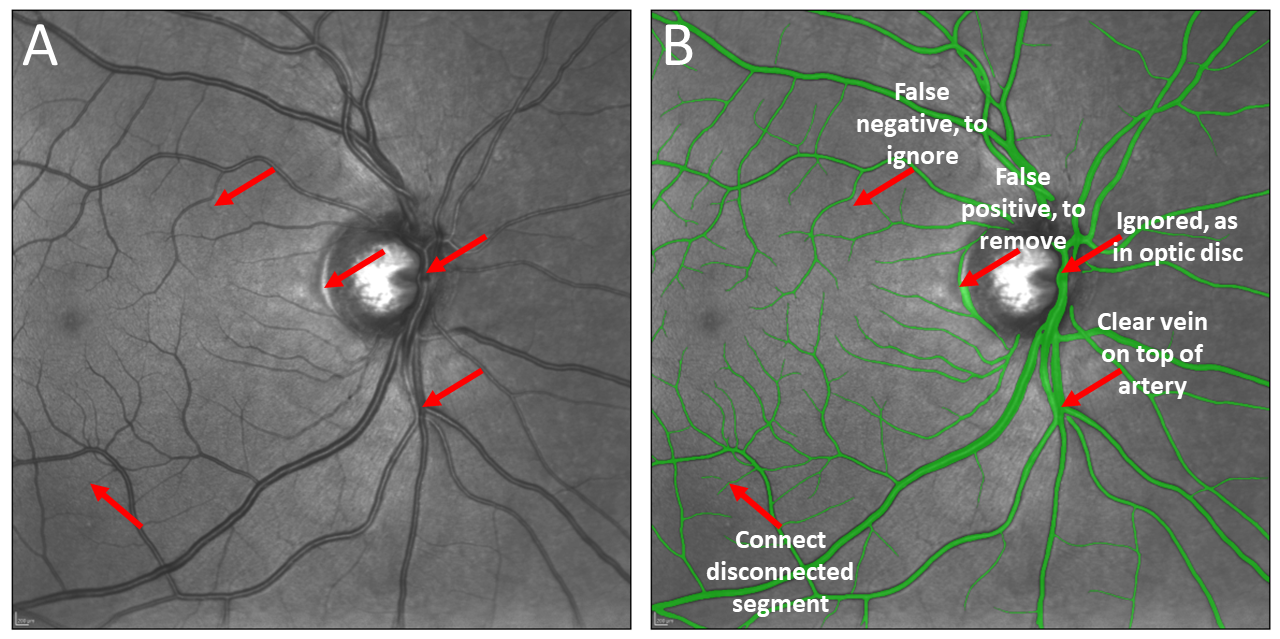}
\caption{An exemplary SLO image (A) with a raw binary vessel mask (B) superimposed in green with annotations to describe each segmentation rule followed during annotation.}
\label{suppfig:AV_protocol} 
\end{figure}

Supplementary \cref{suppfig:AV_protocol} shows an exemplary SLO image with a binary vessel map overlaid using the binary vessel detection model, with annotations to highlight each of the four segmentation rules each grader followed.

\section*{Inter-rater agreement of graders for artery-vein-optic disc detection}
Supplementary \cref{supptab:avod_interrater} shows the results of comparing the segmentations for each task (vessel/artery/vein/optic disc) across the three raters using the dice coefficient (ignoring background). All dice scores were greater than 0.93, suggesting strong agreement between graders.
%Supplementary \cref{suppfig:interrater_disagree} show the greatest disagreements between each pairwise set of graders.

\begin{table}[H]
\centering
{
\begin{tabular}{lllll}
\toprule
Pairwise grader & Vessel & Artery & Vein   & Optic disc \\
\midrule
Grader 1 vs. Grader 2       & 0.99 & 0.96 & 0.97 & 0.94     \\
Grader 1 vs. Grader 3       & 0.99 & 0.95 & 0.96 & 0.95     \\
Grader 2 vs. Grader 3       & 0.99 & 0.95 & 0.96 & 0.94     \\
\bottomrule
\end{tabular}}
\caption{Inter-grader agreement between the three segmentation raters using the Dice  similarity coefficient.}
\label{supptab:avod_interrater} 
\end{table}

\section*{Qualitative evaluation of artery and vein ground truth labels}
Supplementary \cref{supptab:qualitative_segs} show the qualitative grading results from the clinical ophthalmologist (I.M.) for the manual annotations. All SLO images were graded as having `good quality' or above, except for one, and there was no annotation which received a bad or worse rating. Across the image quality categories, a total of 19 SLO images were marked as `very good', 10 as `good' and 1 as `okay'.

The SLO image and manual annotation which had the poorest score is shown in supplementary \cref{suppfig:poorest_adjud}. The clinical ophthalmologist identified an arteriole and venule nasal to the optic disc as misclassified (red arrows). This error amounted to a drop in score from `good' to `okay', suggesting that all other images which rated as `good' had significantly smaller miss-classifications, if any.

\begin{table}[H]
\centering
{\small
\centerline{\begin{tabular}{@{}lc@{}}
\toprule
SLO Image quality & Artery-vein rating \\ \midrule
\multicolumn{1}{l}{Very good ($n=10$)} & VG: 7, G: 3, O: 0, B: 0, VB: 0 \\
\multicolumn{1}{l}{Good ($n=13$)} & VG: 7, G: 5, O: 1, B: 0, VB: 0  \\
\multicolumn{1}{l}{Okay ($n=7$)} & VG: 5, G: 2, O: 0, B: 0, VB: 0\\ 
\bottomrule
\end{tabular}}}
\caption{Qualitative adjudication of the manual annotations for artery and vein classification from a clinical ophthalmologist (I.M.). VG, very good; G, good; O, okay; B, bad; VB, very bad.}
\label{supptab:qualitative_segs} 
\end{table}

\begin{figure}[H]
\centering
\includegraphics[width=0.7\textwidth]{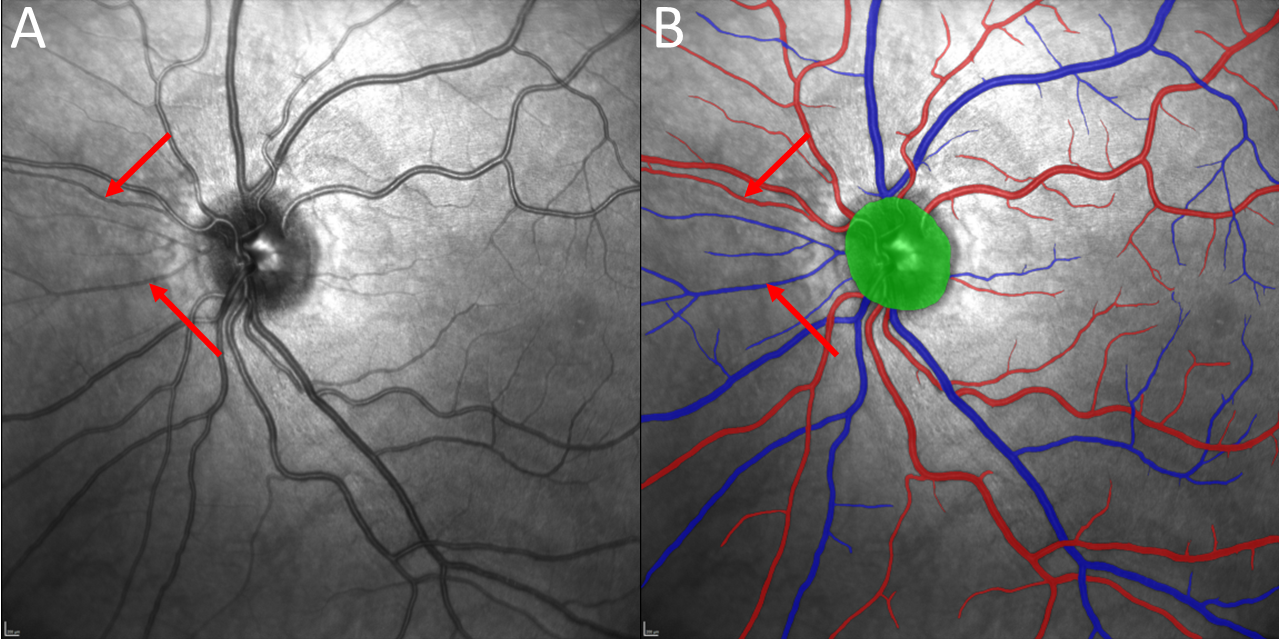}
\caption{SLO image (A) and corresponding manual annotation superimposed (B) with poorest score from manual adjudication from clinical ophthalmologist (I.M.). The red arrows indicate where an arteriole/venule was miss-classified.}
\label{suppfig:poorest_adjud} 
\end{figure}

\section*{RAVIR segmentation label corrections}
During the construction of the artery-vein-optic disc detection (AVOD) model, we intended on using the publicly available RAVIR dataset to further enhance and diversify our dataset of SLO images. However, during experimentation, we observed some inconsistent artery and vein classification. 

\begin{figure}[H]
\centering
\includegraphics[width=0.9\textwidth]{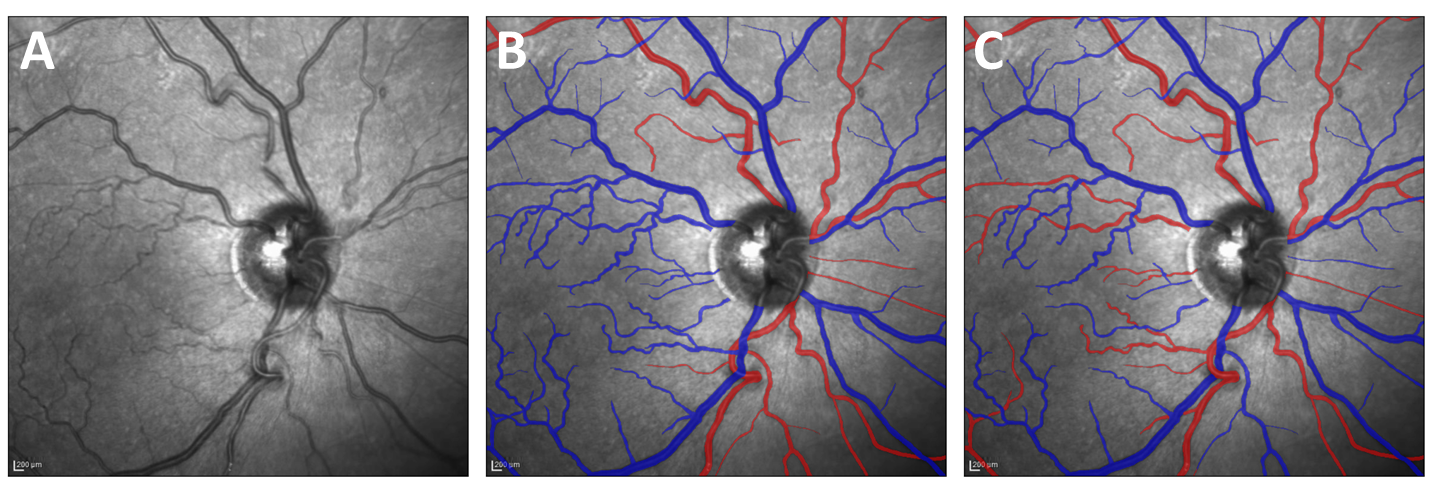}
\caption{SLO image (A), original manual annotation superimposed \cite{hatamizadeh2022ravir} (B) and corrected manual annotation (C). Here, many of the vessels in the macula were all assumed to be venules in the original annotation.}
\label{suppfig:ravir_correction} 
\end{figure}

One of our image graders (author J.B.) identified any major errors in the original segmentation labels and corrected them using ITK-Snap \cite{py06nimg}. Supplementary \cref{suppfig:ravir_correction} shows an SLO image with the most significant error (A), the SLO image with the original labelling overlaid (B), and the SLO image with the corrected version overlaid (C). 

After major corrections were made to all SLO images, we asked a clinical ophthalmologist (author I.M.) to rate the quality of the SLO image, as well as the segmentations before and after correction in a masked and randomised fashion using a 5-point ordinal scale from 2 (very good) to -2 (very bad). We also asked him to rate his preference, with options for both or neither available. Supplementary \cref{supptab:ravir_adjud} shows the results from this adjudication, stratified by image quality.

\begin{table}[H]
\centering
{
\centerline{\begin{tabular}{@{}lccc@{}}
\toprule
Image Quality & Original \cite{hatamizadeh2022ravir} & Corrected & Both \\
\midrule
Very good ($N=6$) & 0 & \underline{3} & \underline{3}\\ 
Good ($N=4$) & 0 & \underline{2} & \underline{2} \\ 
Okay ($N=10$) & 0 & \textbf{7} & 3\\ 
Bad ($N=3$) & 0 & \textbf{2} & 1 \\ 
\midrule
Total ($N=23$) & 0 & \textbf{14} & 9\\
\bottomrule
& & & \\
\toprule
Label & \multicolumn{2}{c}{Artery-Vein classification} & \\
\midrule
Original \cite{hatamizadeh2022ravir} & \multicolumn{2}{c}{VG: 6, G: 8, O: 6, B: 2, VB: 1} & \\
Corrected & \multicolumn{2}{c}{\textbf{VG: 10, G: 11, O: 2, B: 0, VB: 0}} &\\ 
\bottomrule
% \hspace{0.5em}
\end{tabular}}}
\caption{Qualitative adjudication between the original and corrected manual annotations for artery and vein classification in the RAVIR dataset \cite{hatamizadeh2022ravir}, from a clinical ophthalmologist (I.M.). VG, very good; G, good; O, okay; B, bad; VB, very bad.}
\label{supptab:ravir_adjud} 
\end{table}

We chose to exclude the RAVIR dataset from training the artery-vein-optic disc model and used the corrected segmentation labels as an external test set for our model. We hypothesised this set would pose a significant challenge for our model, which was trained on images related to systemic health.

\section*{SLOctolyzer's segmentation model architectures}

\subsection*{Binary vessel detection}
For the binary, all-vessel segmentation model, we use a custom UNet deep learning architecture with a depth of 4, as shown in supplementary \cref{suppfig:binary_Unet}. A convolution block consists of two $3 \times 3$ convolution layers, each followed by batch normalisation \cite{ioffe2015batch} and non-linear rectified linear unit activation (ReLU). After an initial convolution block, the model's DownBlocks double the channel dimension from 48 to 768 while decreasing the spatial dimension by $\nicefrac{1}{2}$ to a final, encoded block with a spatial resolution of 20 $\times$ 15 with 768 features maps. The model's UpBlocks reverse the DownBlocks, with each block reducing the channel dimension by $\nicefrac{1}{2}$ after applying a $2 \times 2$ transposed convolution layer with a $2 \times 2$ stride which scales the spatial dimension by 2. A final $1 \times 1$ output convolution layer reduces the channel dimension to 1, resulting in the probabilistic segmentation map as output 

As described in the main text, offline augmentation extracts 20 random patches of size 320 $\times$ 240 pixels from each SLO image (of native image resolution 768 $\times$ 768 pixels) and these are used for training, as shown in the input/output stage of supplementary \cref{suppfig:binary_Unet}. At inference, the entire SLO image of image resolution 768 $\times$ 768 pixels is fed into the network, outputting a 768 $\times$ 768 pixel probability segmentation map.

\begin{figure}[H]
\centering
\includegraphics[width=0.9\textwidth]{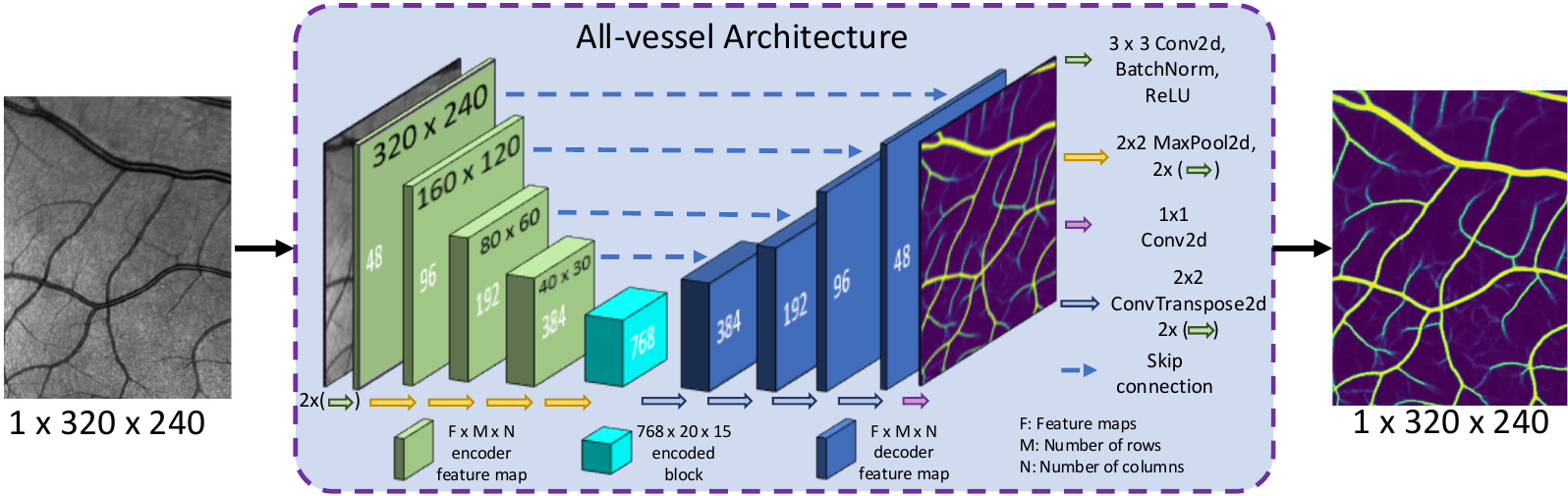}
\caption{Custom UNet model architecture for the binary, all-vessel segmentation model.}
\label{suppfig:binary_Unet} 
\end{figure}
\clearpage
\subsection*{Fovea detection}
The model used for fovea detection was an `off-the-shelf', pre-trained model from SegmentationModelsPytorch \cite{Iakubovskii2019} which utilised a MobileNetV3 \cite{howard2019searching} backbone encoder arm with a UNet decoder arm (the default setup from SegmentationModelsPytorch). The exact Python code to instantiate the model is seen below, with the SegmentationModelsPytorch package assumed to be installed.
\begin{mypython}[caption={Python code to reproduce model for fovea detection.},label=suppcode:fovea_code]
import segmentation_models_pytorch as smp
model = smp.Unet(encoder_name="timm-mobilenetv3_small_075", 
                 encoder_weights="imagenet", 
                 in_channels=1, 
                 classes=1)
\end{mypython}

\subsection*{Artery-vein-optic disc detection}
The model used for artery-vein-optic disc detection was an `off-the-shelf', pre-trained model from SegmentationModelsPytorch \cite{Iakubovskii2019} which utilised a ResNet101 \cite{he2016deep} backbone encoder arm with a UNet decoder arm (the default setup from SegmentationModelsPytorch). The exact Python code to instantiate the model is seen below, with the SegmentationModelsPytorch package assumed to be installed.
\begin{mypython}[caption={Python code to reproduce model for artery-vein-optic disc detection.},label=suppcode:avod_code]
import segmentation_models_pytorch as smp
model = smp.Unet(encoder_name="resnet101", 
                         encoder_weights="imagenet", 
                         in_channels=1, 
                         classes=4)
\end{mypython}

\section*{Global and local vessel calibre}
Measurements of fractal dimension, vessel density and global vessel calibre are not taken across zones B and C because they capture a global measure of the vasculature. Instead, for zones B and C, we compute tortuosity density, central retinal artery/vein equivalents, and local vessel calibre along individual vessel segments in the vessel maps. These vessel segments are defined where crossings and bifurcations are observed.

\begin{figure}[H]
\centering
\includegraphics[width=0.9\textwidth]{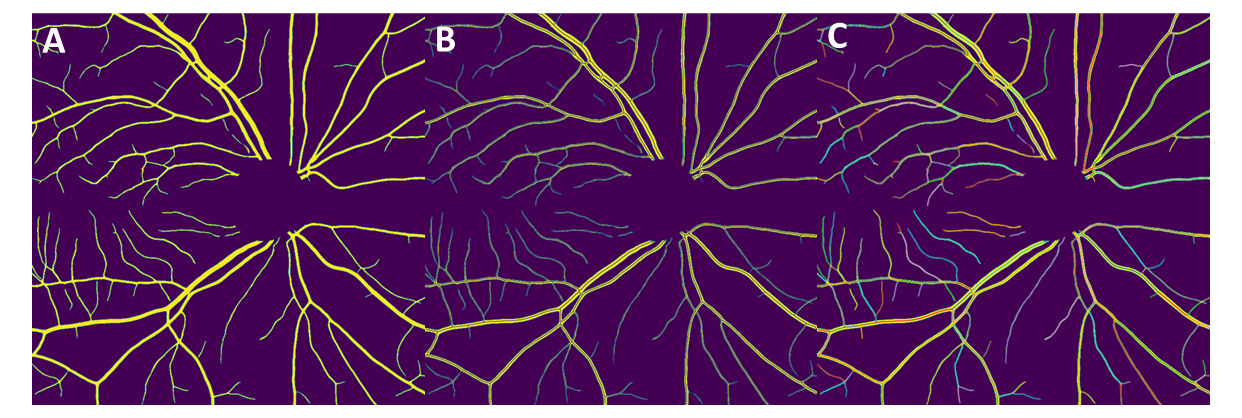}
\caption{(A) SLO binary vessel map. Binary vessel map with skeleton overlaid in brown (B) and individual vessel segments overlaid in an array of colours (C). Note that the optic disc area is removed from the vessel map during feature measurement.}
\label{suppfig:global_vs_local_calibre} 
\end{figure}

Supplementary \cref{suppfig:global_vs_local_calibre} shows the key differences in global and local vessel calibre. Global vessel calibre is the ratio between vessel pixels and skeletonised vessel pixels (number of yellow pixels divided by brown pixels). This provides a coarse measure of calibre. Local vessel calibre measures the diameter of each vessel segment by traversing it's length and measuring the diameter of the largest circle which fits entirely within it. A vessel segments' calibre is then the average value of the list of circle diameters collected as the vessel segment is traversed. Local vessel calibre is thus the overall average of all vessel segment calibres, hence a more granular approach to assess vessel calibre.

\section*{SLOctolyzer's interface}

\begin{figure}[H]
    \centering
    \includegraphics[width=0.85\textwidth]{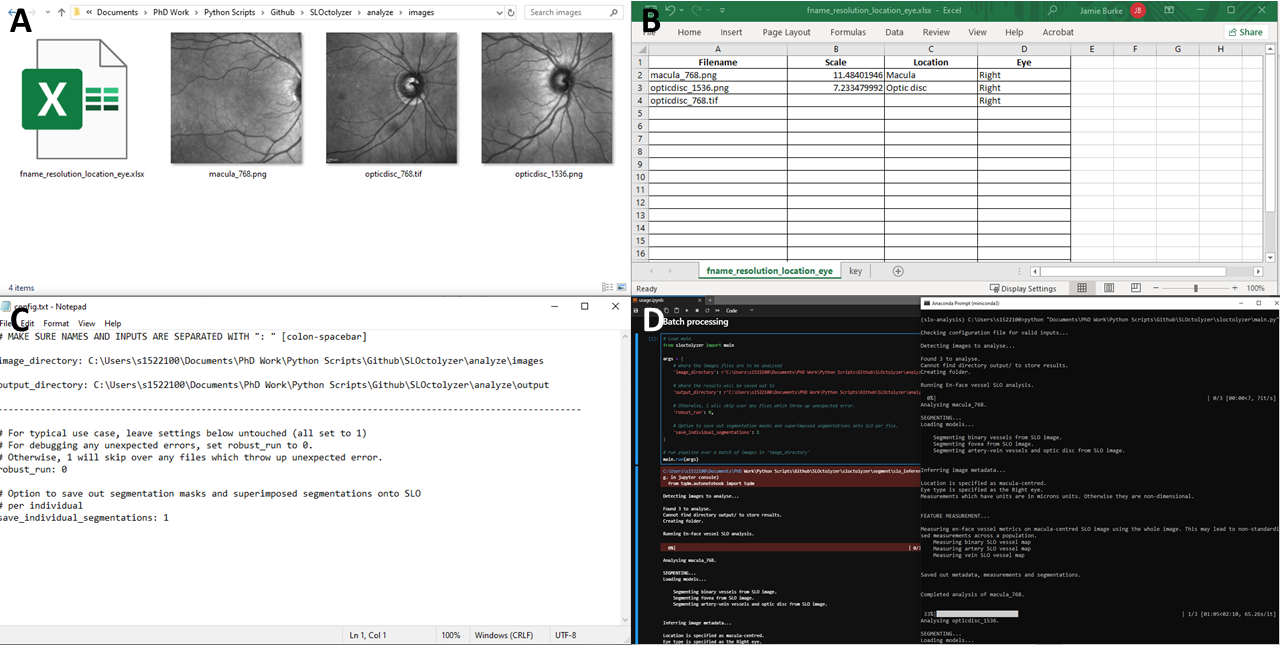}   
    \caption{Summary of the process of running SLOctolyzer on a batch ($n=3$) of images. (A) Directory of input SLO images to be analysed, with an optional spreadsheet which allows manual data entry of transversal spatial sampling length-scale, laterality and location. (B) Spreadsheet with optional metadata to accompany SLO images during pipeline. Note that missing information is supported. (C) Configuration file specifying the file paths to images and where results will be saved. (D) Demonstration of running SLOctolyzer via JupyterLab or via the terminal.}
    \label{suppfig:sloctolyzer_input}
\end{figure}

Supplementary \cref{suppfig:sloctolyzer_input} shows the core steps to run the SLOctolyzer pipeline. SLOctolyzer can be run either from an integrated development environment (IDE), such as Jupyter, Spyder or Visual Studio Code, or directly from the terminal. All that is required is the input/output directories in a simple configuration text file. A process log for the user is outputted as the pipeline processes each image in turn, utilising GPU acceleration if available. When inputting image files, an optional spreadsheet with additional metadata such as the eye laterality, location and transversal spatial sampling length-scale (in microns-per-pixel) can be provided, so that measurements in pixel units can be converted into microns. For \verb|.vol| this is not necessary as SLOctolyzer can access the file metadata.

\begin{figure}[H]
    \centering
    \includegraphics[width=0.85\textwidth]{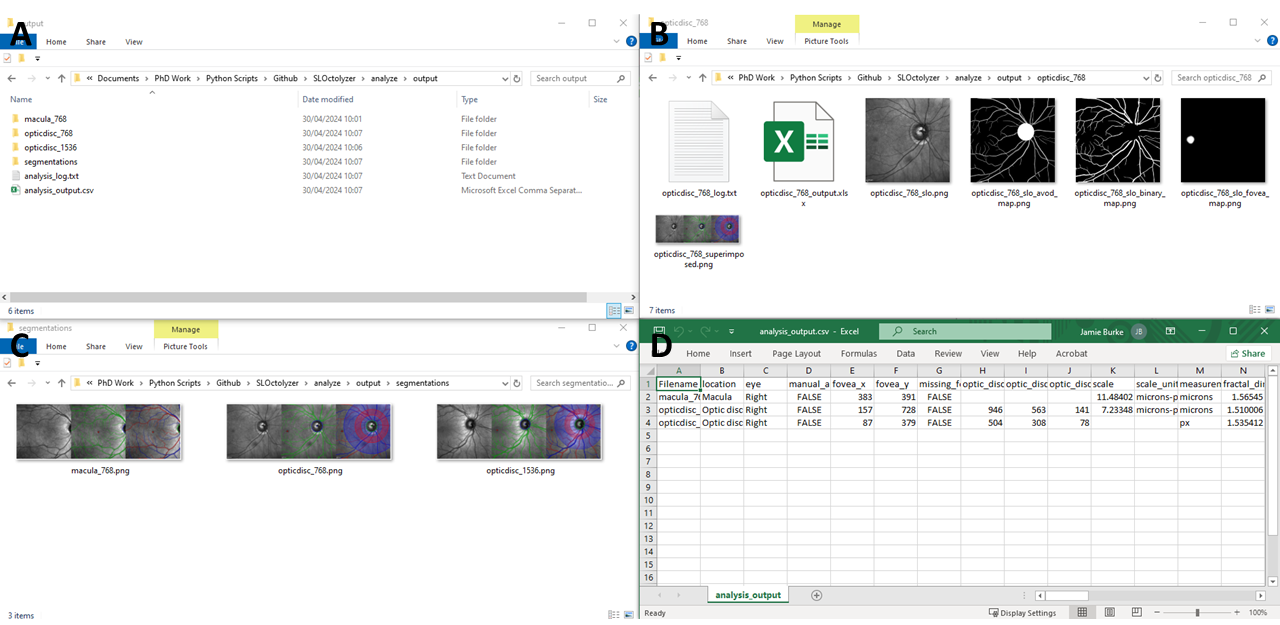}   \caption{Summary of the output of running SLOctolyzer on a batch ($n=3$) of images. (A) Directory of output folders for each of the three image files analysed. (B) Output files for the `opticdisc\_768' exemplar SLO image, including the original SLO image, segmentation masks, composite image with segmentations superimposed and a results file with metadata and feature measurements. (C) A directory with composite images of SLO and segmentation masks per image file for quick inspection of segmentation quality. (D) A composite results file with the metadata and features of each image file processed, stored row-wise to facilitate downstream statistical analysis.}
    \label{suppfig:sloctolyzer_output}
\end{figure}

Supplementary \cref{suppfig:sloctolyzer_output} summarises the output from running SLOctolyzer on a batch ($n=3$) of images. A folder is generated for each SLO image file analysed, and within each folder contains the segmentations masks, feature measurements and a composite image showing the segmentations superimposed onto the SLO. For ease of assessing segmentation quality, composite segmentation images are saved out for every SLO image analysed into a separate folder. Finally, an output spreadsheet collates all feature measurements for every image analysed row-wise, which can be quickly loaded into a statistical programming language for data analysis.

\section*{Examples of poor performance from reproducibility analysis}
\begin{figure}[H]
\centering
\includegraphics[width=0.8\textwidth]{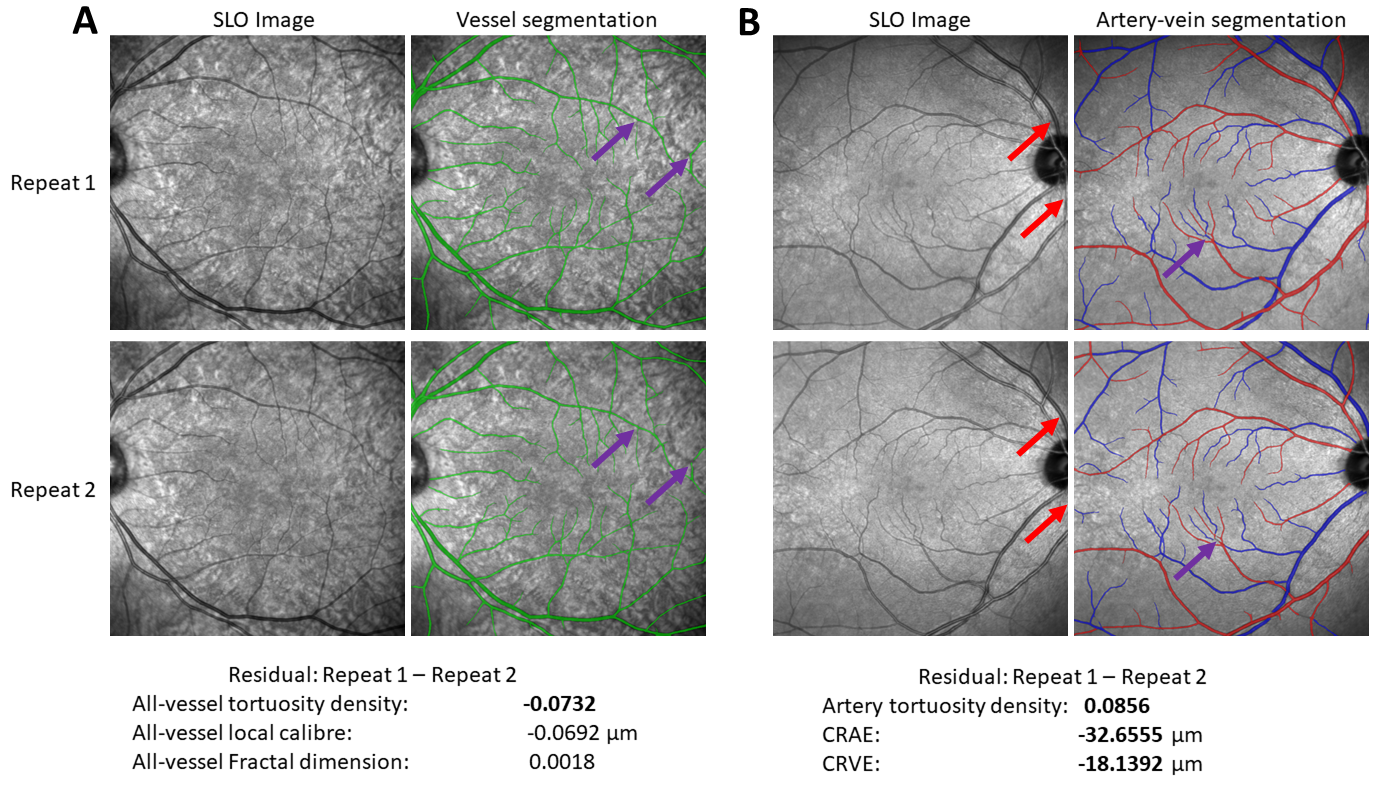}
\caption{Some examples of poor agreement in repeated SLO images for tortuosity density (A), and vessel width measurements (B). In \cref{fig:ind_level_repr}, (A) is represented by the blue star, while (B) is represented by the red stars.}
\label{suppfig:poor_repeatability} 
\end{figure}

Supplementary \cref{suppfig:poor_repeatability} shows two examples which had the second largest outliers for tortuosity density and large-vessel calibre (AVR/CRAE/CRVE). In supplementary \cref{suppfig:poor_repeatability}(A), while there is no qualitative difference in the binary vessel segmentation, the error in tortuosity density is quite large. This is likely due to vessel disconnectedness between the segmentations (purple arrows). Nevertheless, the local vessel calibre and fractal dimension residuals are very low.

Additionally, due to the unregistered nature of the SLO image pairs, minor translations/scales can have an inordinate impact on large-vessel calibre reproducibility given the major arteries and veins appear toward the edge of macula-centred SLO images. In supplementary \cref{suppfig:poor_repeatability}(B), this particular repeated set of SLO images are off by a horizontal translation while intersecting the disc and thus cuts off one of the major arteries and veins (red arrows). This had a significant consequence to the CRVE and CRAE residuals, causing large error. Nevertheless, local vessel calibre remains stable due to taking into account the whole vasculature of the image. 

\end{document}